\documentclass{article}

\usepackage{arxiv}

\usepackage[utf8]{inputenc} 
\usepackage[T1]{fontenc}    
\usepackage{hyperref}       
\usepackage{url}            
\usepackage{booktabs}       
\usepackage{amsfonts}       
\usepackage{nicefrac}       
\usepackage{microtype}      
\usepackage{lipsum}		
\usepackage{graphicx}
\usepackage{natbib}
\usepackage{doi}
\usepackage[most]{tcolorbox}
\usepackage{amsmath,amsthm}

\newtheorem{theorem}{Theorem}
\newtheorem{lemma}[theorem]{Lemma}      

\newtheorem{proposition}[theorem]{Proposition}

\theoremstyle{definition}

\theoremstyle{remark}
\newtheorem*{remark}{Remark}            

\title{Combining Adaptive Seamless Phase II/III Designs and Group Sequential Analysis for Time-to-Event Endpoints in Oncology Trials}

\author{ \href{https://orcid.org/0000-0000-0000-0000}{\includegraphics[scale=0.06]{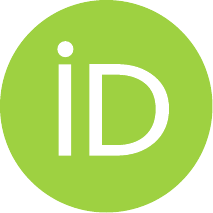}\hspace{1mm}Wentian Guo}
  \\
  AstraZeneca\\
  \texttt{wentian.guo@astrazeneca.com} \\
  \AND
  Qingyu Chen \\
  FDA \\
  \texttt{Qingyu.Chen@fda.hhs.gov} \\
  \And
  Chi Song \\
  Amgen \\
  \texttt{csong06@amgen.com} \\
  \And
  Jiabu Ye \\
  AstraZeneca \\
  \texttt{jiabu.ye1@astrazeneca.com} \\
  \AND
  Chintu Desai \\
  AstraZeneca \\
  \texttt{chintu.desai@astrazeneca.com} \\
  \AND
  Victor Zhang \\
  AstraZeneca \\
  \texttt{victor.zhang1@astrazeneca.com} \\
  \And
  Rui Miao \\
  AstraZeneca \\
  \texttt{rui.miao1@astrazeneca.com} \\
}

\renewcommand{\headeright}{}
\renewcommand{\undertitle}{}
\renewcommand{\shorttitle}{}
\providecommand{\triangleeq}{\mathrel{\stackrel{\triangle}{=}}}

\hypersetup{
pdftitle={Appendix: multiplicity control of OS},
pdfsubject={q-bio.NC, q-bio.QM},
pdfauthor={Wentian Guo, Chi Song, Jiabu Ye, Victor Zhang, Rui Miao},
pdfkeywords={seamless Phase II/III design, dose optimization, combination p-value statistics, closed testing procedure, group sequential analysis, multiplicity control},
}

\begin{document}
\maketitle

\begin{abstract}
Seamless Phase II/III designs are a natural vehicle for the randomized dose
optimization called for by the US Food and Drug Administration's Project Optimus
initiative. Their extension to group sequential repeated testing of a
time-to-event endpoint after end-of-Phase-II dose selection, however, remains
largely unestablished. Existing proposals index the boundaries for the
post-selection looks by the overall information fraction of the primary
endpoint---the correct index for the log-rank statistics of an unadapted trial,
but not for the combination statistics these designs actually test---and support
family-wise error rate (FWER) control by simulation. We instead work from the
joint distribution of the combination statistics themselves, calibrating group
sequential boundaries separately for each hypothesis in the closure and applying
closed testing at every look; a general lemma establishes that this controls the
FWER in the strong sense, for any number of hypotheses and analysis times.
Specifically, under patient-wise separation and an inverse-normal combination
test, the interim and final statistics for an individual hypothesis are bivariate
normal, with a correlation determined by how the events at each look divide
between the pre- and post-selection cohorts rather than by the overall
information fraction, so that their boundaries follow from standard group
sequential software; the corresponding statistics for the intersection
hypothesis, formed from a Hochberg-adjusted Stage-1 $p$-value, are not
multivariate normal, and we obtain their boundaries by a dedicated numerical
integration. In a simulation study based on a hypothetical oncology trial, the
procedure held the one-sided FWER at 2.3\%--2.4\% against a 2.5\% nominal level
and attained 83\%--92\% power under equally-effective alternatives, 8--10
percentage points above an operational seamless design that discards Stage~1
data. Head-to-head against a recently proposed incremental-Dunnett procedure, the
two were near-identical when the design projections were correct, but ours was more robust when the combination weights were based on a misprojected division of the Stage~1 events either side of the interim.
\end{abstract}


\section{Introduction}
As with the Project Optimus initiative by the US Food and Drug Administration (FDA), it has become an obligation for the pharmaceutical companies to optimize the dose in oncology clinical trials. As pointed out by \citet{fourie2022improving}, the strategy of dose optimization in oncology drug development includes considering randomized evaluation of multiple dosages in a trial aiming to establish superiority of the investigational drug to a standard-of-care comparator. An efficient way to do so might be an adaptive seamless phase II/III trial design, which starts with multiple treatment arms and drops sub-optimal arm(s) during the trial (here a lot of references for seamless design). Seamless design can be generally classified into operational seamless design and inferential seamless design. In this paper we will focus on inferential seamless design for the statistical efficiency it brings.\\
\\
Various methods for controlling the FWER in seamless phase II/III trial design with dose selection have been proposed by different authors, including the group-sequential methods \citep{stallard2003sequential,stallard2011group}, combination p-value methods in conjunction with closed testing procedure (CTP) \citep{bauer1999combining,bretz2006confirmatory}, the conditional error methods in conjunction with CTP \citep{koenig2008adaptive}, and the graphical method \citep{sugitani2016simple}. The graphical method is easy to implement, although coupled with conservative performance in terms of power due to the Bonferroni method base.  \\
\\
However, there exists an extra layer of complexity when applying the seamless phase II/III design to a trial with time-to-event endpoint. As the events from the subjects enrolled before the dose selection cannot be fully observed at the time of dose selection, it is usually the case that the dose selection will base on the totality of evidence observed including information of short-term surrogate endpoint. However, as pointed out firstly by \citet{bauer2004modification}, although in the context of sample size re-estimation, using information correlated with the outcome of patients ‘surviving’ the interim analysis would compromise the type 1 error rate even when using combination test statistics (unless the correlation structure is known and appropriately corrected for). This may introduce the dependency between the increment of the log-rank statistics and the selected dose if information of the subjected enrolled before the selection is also included in the increment, that is, the independent increment of the log-rank statistics which is assumed in group sequential methods no longer holds. \citet{di2011time} extended this discussion of type 1 error inflation to the context of the conditional error function (CEF) approach proposed by \citet{koenig2008adaptive} when one tries to apply this approach to TTE endpoint, and pointed out that the FWER may not be controlled when the dose selection also leverages the information of surrogate endpoints at the interim analysis and the correlation between the surrogate endpoint and primary endpoint can not be correctly incorporated. In addition, if one were to use the class of group sequential methods, the correlation of the surrogate endpoint and primary endpoint needs also be correctly incorporated to pertain the FWER control as pointed out by \citet{stallard2011group}. It is, however, difficult to account for the correlation of the surrogate endpoint and primary endpoint at the design stage though, as discussed in many academic articles \citep[see, e.g.,][]{belin2020progression} as well as regulatory agencies' guidance.\\
\\
For the methods based on combined stage-wise p-values, some authors have proposed the so-called stage-wise separation methods to address the potential FWER inflation issue \citep{jenkins2011adaptive,di2011time,carreras2015adaptive}. Generally speaking, including the outcome of patients ‘surviving’ the interim analysis in the stage 1 p-value calculation still preserves FWER control. 
\citet{jorgens2019nested}
proposed another separation approach based on combined stage-wise p-values to enable the early testing of the null H at the time of dose selection by further dividing events before and after interim analysis for the stage 1 patient to two p-values. 
However, all these proposed methods focused on two-stage designs, that is, only one interim with dose selection (or early stopping for efficacy at the same time in \citet{jorgens2019nested}) followed by a final analysis.
In real-world trial design, it may be of great interest to add an interim analysis for potentially rejecting the null H earlier after the dose is selected, and prior to the final analysis. Two proposals try to extend the two-stage adaptive design in exactly this direction: \citet{miao2023gated} with subpopulation selection at the adaptation interim, and \citet{wang2023rank} with treatment selection. 
Both proposals adopt combination $p$-value methods at
all the looks following the adaptation interim, and derive the boundaries for those repeated tests from an off-the-shelf O'Brien--Fleming $\alpha$-spending function indexed by the overall information fraction of the primary endpoint. That indexing is the correct one for the canonical joint distribution, in which the successive log-rank statistics are those of an unadapted trial and their correlation is the square root of the information fraction. It is not the correct one for the combination statistics that these designs actually test. 
Neither paper derives the correlation of the combination statistics, and in both the FWER is supported by simulation. In this paper, we propose a method for such a design based on combination $p$-value statistics, with dose selection at the first interim analysis and potential early stopping at the second interim analysis. Rather than demonstrating FWER control by simulation, we consider the joint distribution of the combination statistics themselves, and derive from it the rejection probability at the interim and final analyses, on which the construction of the boundaries is then based.
\\
\\
A closely related contribution is \citet{zhang2025mams}, who propose a multi-arm multi-stage (MAMS) group sequential Phase 2/3 design with dose selection and a time-to-event endpoint. Their approach shares the same core ingredients as ours: the inverse-normal combination test to combine stage-wise log-rank statistics before and after dose selection, and a closed testing procedure to control multiplicity from comparing several experimental doses to a common control, together with group-sequential $\alpha$-spending for repeated testing in the Phase 3 portion. Compared to \citet{zhang2025mams}, our work differs in two ways. First, their multiplicity adjustment for the intersection hypothesis uses a Dunnett correction, exploiting the joint multivariate-normal distribution of the arm-vs-control log-rank statistics under the shared-control correlation, whereas we use a Hochberg-type correction \citep{sarkar1997simes}, which does not require the full joint distribution.
Second, their construction of the Stage-1 (cohort-1) $p$-value is itself a nested combination test: they decompose the cohort-1 cumulative log-rank statistic into stage-wise \emph{incremental} log-rank contributions across the Phase 3 looks, apply a Dunnett correction to the stage-wise incremental statistics separately at each look, and then combine the resulting stage-wise Dunnett $p$-values via an inner inverse-normal combination test. Our construction, in contrast, uses a single Stage-1 $p$-value based on the cumulative log-rank statistic on cohort-1 patients at each look (with patient-wise separation) and applies the Hochberg correction only once, without further decomposing the Stage-1 statistic across looks.
\\
\\
This paper is laid out as below. Section \ref{case} will introduce an oncology trial and set the scene for the paper. Section \ref{method} will provide the detailed combination approach to address the multiplicity problem in this type of seamless trial with repeated testing. Section \ref{Simulation Studies} will investigate the operating characteristics of the proposed method and compare it to other types of multiplicity control methods and design strategies. Section \ref{summary} will conclude this paper with summaries and discussion.


\section{Case Study}\label{case}
Consider an inferential seamless phase II/III study starting with two treatment arms and one control arm, and $N_1$ patients being equally randomized to these three arms in the phase II part. At the end of phase II part (or stage 1), one of the treatment arms will be selected to proceed to the phase III part (or stage 2) while the other treatment arm will be dropped. $N_2$ patients will be equally randomized to two arms, the selected treatment arm and the control arm, in the phase III part. After the end of the enrollment, there will be two looks for overall survival (OS) efficacy comparing the selected arm versus the control arm, interim analysis (IA) and final analysis (FA).\\
\\
Follow-up of patients enrolled during the phase II part continues after the dose selection until each analysis time, so that their events also contribute to the phase III analysis; this is why we call the design \emph{inferential} seamless. In practice, it takes some time to prepare the data and make the dose selection decision, so patients continue to be randomized to all three arms until that decision is made. Patients enrolled during this window and assigned to the selected experimental arm or the control arm are also carried into the phase III analysis, while patients assigned to the unselected experimental arm during this window are excluded from the primary analysis. The formal definitions of Stage 1 and Stage 2 patients, along with the patient-wise separation used to construct stage-wise $p$-values, are given in Section~\ref{sec:setup}.\\
\\
The design entails three pre-planned analysis time points, each with a distinct objective:
\begin{enumerate}
    \item Dose selection, based on the totality of efficacy (OS, progression-free survival [PFS], objective response rate [ORR], etc.), safety, pharmacokinetics, and biomarker data. No formal testing of OS efficacy is performed at this time point due to the small sample size and immaturity of events.
    \item Interim analysis (IA), testing OS efficacy with the possibility of stopping early for efficacy.
    \item Final analysis (FA), testing OS efficacy.
\end{enumerate}
The first time point above is sometimes also referred to as an ``interim analysis'' in the seamless-design literature. In this paper, however, we reserve the terms ``interim analysis'' (IA) and ``final analysis'' (FA) exclusively for the two efficacy looks during phase III that carry the potential to reject the null hypothesis; the first time point we refer to simply as ``dose selection.''

\section{Methodology}\label{method}

\subsection{Group sequential approach for repeated log-rank tests}\label{sec:gsd}

Before turning to the seamless-plus-repeated-testing problem, we recall the
canonical two-look group-sequential log-rank test for a two-arm 1:1
randomized controlled trial with primary endpoint OS. At the $l$-th look
($l=1,2$; interim and final) with $S_l$ ordered distinct event times $t_s$
($s=1,\dots,S_l$), the standardized log-rank statistic comparing treatment
$i=1$ to control $i=0$ is
\[
Z_l = \frac{\sum_{s=1}^{S_l}\!\left[d_{1s}-\dfrac{r_{1s}}{r_s}d_s\right]}
          {\sqrt{\sum_{s=1}^{S_l}\dfrac{r_{1s}r_{0s}d_s(r_s-d_s)}{r_s^2(r_s-1)}}},
\]
where $r_{is}$ and $d_{is}$ are, respectively, the number at risk and the
number of events on arm $i$ at $t_s$, $r_s=r_{1s}+r_{0s}$ and
$d_s=d_{1s}+d_{0s}$. Under a hypothesis $H_0:\theta=\theta_0$ on the hazard
ratio, $(Z_1,Z_2)$ is asymptotically bivariate normal:
\begin{align} \label{dist_gsd}
\begin{pmatrix}Z_1\\ Z_2\end{pmatrix}
\sim N\!\left(
\begin{pmatrix}\log(\theta_0)\sqrt{I_1}\\ \log(\theta_0)\sqrt{I_2}\end{pmatrix},
\begin{pmatrix} 1 & \sqrt{I_1/I_2}\\ \sqrt{I_1/I_2} & 1\end{pmatrix}\right),
\end{align}
with information $I_l = (e_{l,1}+e_{l,0})/4$ and $e_{l,i}=\sum_s d_{is}$.

This fixes the sign convention used throughout the paper. A beneficial
treatment has a hazard ratio below one, so $\log(\theta_0)<0$ and by
(\ref{dist_gsd}) an effective treatment drives $Z_l$ downwards: \emph{small}
values of the test statistic are evidence of benefit, and rejection is
therefore in the direction $Z_l\leq c_l$, with critical values $(c_1,c_2)$
that are negative. Group-sequential software conventionally reports
boundaries for the opposite orientation, in which large values favour the
experimental arm; the boundaries here are the negatives of the reported
values, and numerical boundary values are quoted as magnitudes throughout.

The interim and final critical values $(c_1,c_2)$ are then determined under
the null by
\begin{equation}\label{eq1}
Pr(Z_1 \leq c_1 \mid H_0) = \alpha_1,
\end{equation}
\begin{equation}\label{eq2}
Pr(Z_1 > c_1, Z_2 \leq c_2 \mid H_0) = \alpha - \alpha_1,
\end{equation}
where $\alpha$ is the target one-sided FWER and $\alpha_1<\alpha$ is the
amount spent at the interim. Standard choices for $(c_1,c_2,\alpha_1)$
include the O'Brien--Fleming procedure \citep{obrien1979multiple}, Pocock's
approach \citep{pocock1977group}, and the Lan--DeMets $\alpha$-spending
functions \citep{gordon1983discrete}.

\subsection{Notation, hypotheses, and stage-wise \texorpdfstring{$p$-values}{p-values} under seamless design}\label{sec:setup}

We now specialize to the seamless design of Section~\ref{case}. Two
experimental arms $i=1,2$ are compared to a common control arm $i=0$; let
\[
H_{0,i}:\ \theta_i=1,\quad i=1,2,
\]
where $\theta_i$ is the OS hazard ratio (HR) of arm $i$ versus control. At the
end of Stage~1, one experimental arm $i^{*}\in\{1,2\}$ is selected and
Stage~2 enrols only $i^{*}$ and the control arm. The only null hypothesis
of interest at analysis time is therefore $H_{0,i^{*}}$; by the closed
testing procedure
\citep{marcus1976closed,bretz2006confirmatory,bretz2009adaptive,schmidli2006confirmatory},
$H_{0,i^{*}}$ can be rejected at level $\alpha$ if and only if both the
individual hypothesis $H_{0,i^{*}}$ and the global intersection
$\bigcap_{i=1,2}H_{0,i}$ are rejected at level $\alpha$. In the remainder
of this section we will show how the test statistics will be constructed
for testing OS at each look.

\paragraph{Patient-wise separation and stage-wise $p$-values.}
Stage-1 patients are those enrolled before the dose-selection data cut-off (DCO); Stage-2
patients are those enrolled after the DCO on either arm $i^{*}$ or the
control arm. Under \emph{patient-wise separation}
\citep{jenkins2011adaptive,di2011time,carreras2015adaptive}, Stage-1
patients continue to be followed after the DCO and all their events
(including those occurring after Stage~1 has ended) contribute to the
Stage-1 log-rank statistic, not to the Stage-2 one. This construction
preserves type 1 error control even when surrogate information is used for
dose selection.

At each Phase 3 look $k\in\{IA,FA\}$, three one-sided log-rank
$p$-values are computed:
\begin{itemize}
\item $p_{k,1,i}$ ($i=1,2$): arm $i$ vs.\ control on Stage-1 patients only;
\item $p_{k,2}$: arm $i^{*}$ vs.\ control on Stage-2 patients only.
\end{itemize}
It will be convenient to work with these $p$-values on the normal scale,
and we write
\[
Z_{k,1,i}=\Phi^{-1}(p_{k,1,i}),\qquad \zeta_{k}=\Phi^{-1}(p_{k,2}),
\]
so that $Z_{k,1,i}$ is the Stage-1 log-rank statistic for arm $i$ vs.\
control at look $k$ (three subscripts denoting look, stage and arm, as for
the event counts $e_{k,j,i}$ below) and $\zeta_{k}$ is the corresponding
Stage-2 statistic for the selected arm. Under the relevant null each is
standard normal.

The Hochberg-adjusted Stage-1 $p$-value for the intersection hypothesis
\citep{hochberg1988sharper,dunnett1992stepup,dmitrienko2009multiple} is
\begin{equation}\label{eq:hb}
p_{k,1}^{HB}=\min\bigl\{2\min(p_{k,1,1},p_{k,1,2}),\ \max(p_{k,1,1},p_{k,1,2})\bigr\}.
\end{equation}
After dose selection, the Stage-2 $p$-value for the intersection
hypothesis coincides with that for the individual hypothesis
\citep{bretz2009adaptive}, so both are denoted $p_{k,2}$ without ambiguity.

\paragraph{Combined test statistics.}
For a Stage-1 $p$-value $q$ and pre-specified weights $(w_{k,1},w_{k,2})$
with $w_{k,1}^2+w_{k,2}^2=1$, define the inverse-normal weighted
combination at look $k$ as
\begin{equation}\label{eq:combined}
c_k(q,p_{k,2}) \;=\; w_{k,1}\,\Phi^{-1}(q) + w_{k,2}\,\Phi^{-1}(p_{k,2}).
\end{equation}
Equation~(\ref{eq:combined}) uses $\Phi^{-1}(q)$ while the
combination-test literature conventionally writes $\Phi^{-1}(1-q)$
\citep{bauer1999combining,bretz2009adaptive}. The two forms differ only by
an overall change of sign, and we adopt the former so that the combination
statistics keep the orientation of the log-rank statistics of
Section~\ref{sec:gsd}: small values are evidence of benefit, and every
rejection region below is of the form ``statistic at or below its
boundary''. 

Substituting the two choices of $q$ gives the test statistic for each
hypothesis at look $k$:
\begin{itemize}
\item Individual $H_{0,i^{*}}$: use $q = p_{k,1,i^{*}}$.
\item Intersection $\bigcap_i H_{0,i}$: use $q = p_{k,1}^{HB}$.
\end{itemize}
The weights $(w_{k,1},w_{k,2})$ are pre-specified in the statistical analysis plan (SAP) and may
differ between looks and, in principle, between the individual and
intersection tests. In the simulation study of
Section~\ref{Simulation Studies} we take the same $(w_{k,1},w_{k,2})$
for both the individual and the intersection tests at each look $k$,
and unless otherwise stated we set $w_{k,j}^2$ proportional to the
expected number of Stage-$j$ events from the selected experimental arm
and the control arm combined at look $k$
(i.e.\ $w_{k,j}^2 = E_{k,j}/(E_{k,1}+E_{k,2})$, where $E_{k,j}$ denotes
the expected number of Stage-$j$ events on the selected arm and control
arm combined at look $k$).

\subsection{A FWER-preserving testing procedure for group sequential boundaries nested within CTP}\label{sec:general}

Rejecting at each look requires a boundary for each of the two test
statistics (individual and
intersection) from (\ref{eq:combined}). We adopt the strategy of nesting
group-sequential boundaries within the closed testing procedure, i.e.\
calibrating the boundaries for the individual and intersection
hypotheses \emph{separately}. That this strategy controls the FWER at
level~$\alpha$ in the strong sense --- that is, for every configuration
of true and false hypotheses, not merely under the global null --- holds
generally, for any number of hypotheses and any number of analysis
times, as stated in Lemma~\ref{lem:key} and proved in
Appendix~\ref{app:lemma1}.

For our two-arm two-look case, the remaining task is to specify the
boundaries for each of the two hypotheses at each of the two looks (IA
and FA).
The two
hypotheses share the combination form (\ref{eq:combined}) and differ only
in which Stage-1 $p$-value is fed in.

\paragraph{Individual hypothesis $H_{0,i^{*}}$.}\label{sec2.1}
Setting $q=p_{k,1,i^{*}}$ in (\ref{eq:combined}) gives the test statistics
\begin{equation}\label{eq5}
c_{IA}(p_{IA,1,i^*},p_{IA,2}) \;=\; w_{IA,1}\Phi^{-1}(p_{IA,1,i^*}) + w_{IA,2}\Phi^{-1}(p_{IA,2}),
\end{equation}
\begin{equation}\label{eq6}
c_{FA}(p_{FA,1,i^*},p_{FA,2}) \;=\; w_{FA,1}\Phi^{-1}(p_{FA,1,i^*}) + w_{FA,2}\Phi^{-1}(p_{FA,2}).
\end{equation}
Under $H_{0,i^{*}}$ the pair $(c_{IA},c_{FA})$ is bivariate normal with
mean $0$, unit variance, and covariance
\begin{equation}\label{eq:corr}
    R \;=\; w_{IA,1}w_{FA,1}\sqrt{\frac{e_{IA,1,i^{*}}+e_{IA,1,0}}{e_{FA,1,i^{*}}+e_{FA,1,0}}}
    \;+\;
    w_{IA,2}w_{FA,2}\sqrt{\frac{e_{IA,2,i^{*}}+e_{IA,2,0}}{e_{FA,2,i^{*}}+e_{FA,2,0}}},
\end{equation}
where $e_{k,j,i}$ is the number of events on arm $i$ from Stage-$j$
patients observed at look $k$. Because the joint distribution
is bivariate normal with known covariance, boundaries $Z_{IA}^{*}$ and
$Z_{FA}^{*}$ can be calibrated using standard group-sequential software
under the canonical formulation of Section~\ref{sec:gsd} (with $(c_{IA},
c_{FA})$ playing the role of the log-rank statistics $(Z_1, Z_2)$ and
$R^{2}$ from (\ref{eq:corr}) providing the information fraction), so that
$H_{0,i^{*}}$ is rejected when $c_{IA}\leq Z_{IA}^{*}$ at IA or
$c_{FA}\leq Z_{FA}^{*}$ at FA. The sign convention of
Section~\ref{sec:gsd} carries over: the $p$-values enter
(\ref{eq:combined}) through $\Phi^{-1}$, so small values of $c_k$ are
evidence against the null and $(Z^{*}_{IA},Z^{*}_{FA})$ are negative, as
are $(c_1,c_2)$ in (\ref{eq1})--(\ref{eq2}).

\paragraph{Intersection hypothesis $\bigcap_{i=1,2}H_{0,i}$.}\label{sec:inter}
Setting $q=p_{k,1}^{HB}$ from (\ref{eq:hb}) in (\ref{eq:combined}) gives
the test statistics
\begin{equation}\label{eq7}
c_{IA}(p_{IA,1}^{HB},p_{IA,2}) \;=\; w_{IA,1}\Phi^{-1}(p_{IA,1}^{HB}) + w_{IA,2}\Phi^{-1}(p_{IA,2}),
\end{equation}
\begin{equation}\label{eq8}
c_{FA}(p_{FA,1}^{HB},p_{FA,2}) \;=\; w_{FA,1}\Phi^{-1}(p_{FA,1}^{HB}) + w_{FA,2}\Phi^{-1}(p_{FA,2}).
\end{equation}
Unlike the individual case, the joint distribution of $(c_{IA},c_{FA})$
is not bivariate normal because $p_{k,1}^{HB}$ is a non-linear transform
of the two underlying arm-vs-control statistics. Standard
group-sequential software therefore cannot be used directly, and we must
obtain the corresponding boundaries $\eta$ (IA) and $\gamma$ (FA) by a
dedicated numerical integration, developed in Section~\ref{sec3} below.

\subsection{Boundaries for repeated testing of the intersection hypothesis}\label{sec3}
Denote the critical values of the intersection-hypothesis combination test statistics (\ref{eq7}) and (\ref{eq8}) by $\eta$ (IA) and $\gamma$ (FA). Type 1 error control at the one-sided level $\alpha$ requires
\begin{equation}\label{eq9}
Pr\left(c_{IA}(p_{IA,1}^{HB},p_{IA,2})>\eta,\ c_{FA}(p_{FA,1}^{HB},p_{FA,2})>\gamma\right)\;\geq\;1-\alpha.
\end{equation}
Unlike the individual-hypothesis case, the joint distribution of the two test statistics is not bivariate normal, because the Hochberg-adjusted Stage-1 $p$-value (\ref{eq:hb}) depends on the two arm-vs-control $p$-values only through their order statistics, so that $\Phi^{-1}(p_{k,1}^{HB})$ is a min/max functional of the two Stage-1 statistics rather than a linear combination of them. The left-hand side of (\ref{eq9}) can nonetheless be reduced to a numerically tractable expression by conditioning on the Stage-2 statistics and exploiting the symmetry of the joint distribution of the four Stage-1 statistics $(Z_{IA,1,1},Z_{IA,1,2},Z_{FA,1,1},Z_{FA,1,2})$ under the intersection null. The full derivation is given in Appendix~\ref{app:alphaint}; the result is a function $\alpha(\eta,\gamma)$ evaluable to essentially arbitrary numerical precision by deterministic quadrature. That derivation is specific to two looks. It is not the only route to these boundaries: Appendix~\ref{app:Llook} gives a recursion that evaluates the same type 1 error at any number of looks, at a cost linear in that number.

Given a numerically-computed $\alpha(\eta,\gamma)$, we adopt the following boundary-selection strategy. The IA boundary is fixed at the corresponding individual-hypothesis value,
\[
\eta \;=\; Z_{IA}^{*},
\]
and the FA boundary is initialized at the individual-hypothesis value $Z_{FA}^{*}$ and iterated as follows. If $\alpha(Z_{IA}^{*},Z_{FA}^{*})\leq\alpha$, we adopt $\gamma=Z_{FA}^{*}$; otherwise we decrease $\gamma$ from $Z_{FA}^{*}$ in small steps (e.g.\ $-0.001$) until $\alpha(Z_{IA}^{*},\gamma)\leq\alpha$. This produces the maximum $\gamma$ that preserves the FWER at a level less than $\alpha$ under the calibration procedure.

Whether the iterative step is needed depends on the combination weights and on the realized information fractions. In the simulation setting of Section~\ref{Simulation Studies}, we found that reusing the individual-hypothesis boundaries directly (i.e.\ setting $\eta=Z_{IA}^{*}$ and $\gamma=Z_{FA}^{*}$) preserved the type 1 error at level $\alpha$ across the full range of stage-1 combination weights examined and across the range of realized information fractions; see Section~\ref{sim:alphaint} for details. The iterative $\gamma$-calibration step is therefore not required for the simulation settings considered in this paper, though it remains available for cases in which the direct reuse of individual boundaries fails.

\section{Simulation Studies}\label{Simulation Studies}

We conducted a simulation study to evaluate the operating characteristics of the proposed procedure under a range of hypothesis, dose selection, and design configurations. Section~\ref{sim:design} describes the shared design and analysis framework and the treatment-effect scenarios used to explore both FWER control (under a global null) and power (under several alternatives), including a comparison against the Dunnett-based procedure of \citet{zhang2025mams} (Z\&J) and an operational seamless design that discards Stage 1 data. Section~\ref{sim:results} reports the results under correctly specified combination weights. Section~\ref{sim:edge} then examines robustness to weight misspecification. Section~\ref{sim:alphaint} numerically verifies, across the range of weights and information fractions arising in the simulation, that reusing the individual-hypothesis group-sequential boundaries for the intersection-hypothesis test preserves the family-wise type~1 error rate, so that the iterative $\gamma$-calibration of Section~\ref{sec3} is not required for the designs considered here.

\subsection{Design}\label{sim:design}

The trial design followed the case study of Section~\ref{case}: three parallel arms (control, low-dose, and high-dose), a two-stage group sequential design, and dose selection at the end of Stage~1. Total sample size was 450 patients accruing uniformly over 24 months.
In Stage~1 (calendar months 0--8, 150 patients), patients were randomized 1:1:1 across the three arms.
The dose selection data cut-off (DCO) occurred at month 8; the selection decision was made at month 10, so patients enrolled between months 8 and 10 continued to be randomized 1:1:1. Stage 1 and Stage 2 patients were defined as in Section~\ref{sec:setup}, with the DCO at month 8 serving as the stage boundary; patients enrolled at or after month 8 on the unselected experimental arm were excluded from the primary analysis. During months 10--24 the two remaining arms (the selected experimental arm and the control arm) accrued at a 1:1 ratio.

For each simulated patient, two endpoints, objective response and time to death, were generated. Objective response was a binary indicator drawn independently as $\mathrm{Bernoulli}(p)$ with the arm-specific ORR $p$. Overall survival (OS) time was drawn from a two-component exponential mixture conditional on response status: non-responders drew from $\mathrm{Exp}(\lambda_N)$ and responders from $\mathrm{Exp}(\lambda_R)$. For the control arm, $(\lambda_R,\lambda_N)$ were solved to jointly satisfy (i) a control ORR of $p=0.30$, (ii) a target marginal median OS of 6~months and (iii) a $2\times 2$ dependence structure at anchor time $t_0=6$~months following the binary-surrogate construction of \citet{jorgens2019nested}, with sensitivity $\mathrm{sens}=0.10$ (equivalently, 90\% of control deaths within $t_0$ occur among non-responders). Taking $t_0$ equal to the target median fixes the marginal probability of death by $t_0$ at $0.5$, which together with (i) and (iii) determines the four cells of the $2\times 2$ table; the two hazards then follow from the response-conditional probabilities of death by $t_0$. This yielded control-arm hazards $\lambda_R=0.0304$ and $\lambda_N=0.1716$, corresponding to responder and non-responder median OS of 22.81 and 4.04~months, respectively, and a Spearman rank correlation between ORR and OS of approximately 0.55.

Seven treatment-effect scenarios were evaluated. In all scenarios the control arm was fixed as described above, and the two experimental arms were specified to cover the response-mediated scenario and scenarios in which the survival benefit is not driven by response. In all alternative scenarios considered, OS is generated from a two-component mixture, and therefore the arm-level hazards are non-proportional and no single hazard ratio describes an experimental arm exactly.  We therefore indexed each experimental arm by its \emph{ratio of medians} (RM), the control marginal median OS (6~months) divided by that of the experimental arm.

\textit{Equally-effective scenarios}, in which both experimental arms are drawn from the same distribution:
\begin{itemize}
    \item \textbf{null}: both arms match control; used to assess FWER.
    \item \textbf{A}: a response-mediated benefit in which the treatment increases ORR from 0.30 to 0.574 while leaving the responder and non-responder OS hazards unchanged, giving an experimental median of 10.0~months and hence RM $=0.60$.
    \item \textbf{B}: both subgroup hazards are scaled by 0.60, keeping ORR at 0.30. Scaling both components gives a median of exactly 10.0~months and hence RM $=0.60$.
    \item \textbf{C}: ORR is held at 0.30 and the responder hazard is left unchanged, with the entire benefit acting through non-responders. That is, $\lambda_N$ is scaled by 0.537 (from 0.1716 to 0.0921, lengthening the non-responder median from 4.04 to 7.53~months), which yields an arm median of 10.0~months and hence RM $=0.60$.
\end{itemize}
Scenario A served as the primary planning scenario from which all design-time quantities were derived. Scenarios B and C serve as sensitivity analyses under alternative biological mechanisms in which ORR is uninformative about arm-level OS benefit.

Scenarios A/B/C have exchangeable treatment arms so that the selection rules cannot be distinguished. To better quantify the benefit of using the short-term endpoint to guide the dose selection, we also explore \textit{Differential scenarios}, in which the low dose is strictly weaker than the high dose under a response-mediated mechanism (i.e. lower ORR but same responder and non-responder OS hazards):
\begin{itemize}
    \item \textbf{A1}: the low-dose arm is null (RM $=1.00$) and the high-dose arm has RM $=0.60$.
    \item \textbf{A2}: low-dose RM $=0.80$ and high-dose RM $=0.60$.
    \item \textbf{A3}: low-dose RM $=0.70$ and high-dose RM $=0.60$.
\end{itemize}
using two dose selection rules applied at the month-8 DCO:
\begin{itemize}
    \item \textbf{ORR-based}: select the experimental arm with the higher observed ORR. ORR was assumed to be observed with a two-month lag following enrolment, so this selection at the month-8 DCO used ORR data on patients enrolled by month~6. Ties were broken at random.
    \item \textbf{OS-based}: fit a Cox proportional hazards model of arm~1 vs.\ arm~2 using OS data censored at the month-8 DCO and select the arm with the smaller estimated hazard.
\end{itemize}

For the primary event-driven analysis, the interim analysis (IA) and final analysis (FA) were triggered when 220 and 300 cumulative OS events, respectively, had been observed among patients in the selected experimental arm and the control arm.  Under the primary design scenario the expected calendar times for IA and FA were approximately 25 and 38 months, respectively.

Efficacy testing followed the closed testing procedure with combination $p$-values described in Section~\ref{method}. 
Group-sequential boundaries for the individual hypothesis were derived from Lan--DeMets $\alpha$-spending with an O'Brien--Fleming shape at the one-sided level $\alpha=0.025$. The information fraction consumed by the spending function is $R^{2}$, where $R$ is the correlation (\ref{eq:corr}) between the IA and FA combination statistics; $R$ depends on how the events at each of the two looks divide between the two cohorts, and at the IA the FA split is not yet known. We therefore fixed the IA boundary at design time and deferred all recalibration to the FA, so that no boundary depends on a quantity unavailable when it is applied. Setting the cohort weights to the expected per-stage event shares at each look, $w_{k,1}^{2}=E_{k,1}/(E_{k,1}+E_{k,2})$ as in Section~\ref{sim:results}, gives
\[
R \;=\; \sum_{j=1}^{2} \sqrt{\frac{E_{IA,j}}{D_{IA}}}\sqrt{\frac{E_{FA,j}}{D_{FA}}}\sqrt{\frac{E_{IA,j}}{E_{FA,j}}}
  \;=\; \sum_{j=1}^{2} \frac{E_{IA,j}}{\sqrt{D_{IA}D_{FA}}}
  \;=\; \sqrt{\frac{D_{IA}}{D_{FA}}},
\]
where $D_{IA}=220$ and $D_{FA}=300$ are the event targets at the two looks, so that the planned information fraction is $R^{2}=D_{IA}/D_{FA}=0.733$ whatever the projected cohort split, and the IA boundary is the design constant $Z_{IA}^{*}=2.371$ ($\alpha$ spent $0.0089$). Because Z\&J's outer weights equal ours and its inner FA weight $w_{3,2}=\sqrt{E_{IA,1}/E_{FA,1}}$ makes its cohort-1 covariance term coincide with ours, the same identity holds for Z\&J, and the IA boundary is therefore common to the two procedures and to both selection rules. At the FA, $R$ was recomputed from the realized per-stage event split---which varies across replicates even though the total event count at each look is fixed by design---and $Z_{FA}^{*}$ was re-solved at that correlation holding the $\alpha$ already spent at the IA fixed, so that the nominal two-look spend is exactly $0.025$ in every replicate. Boundaries for the intersection hypothesis were set equal to the individual-hypothesis boundaries, which, as Figure~\ref{fig:alphaweights} confirms, preserves type 1 error control over the weight combinations and realized information fractions arising here.

To contextualise the inferential design, we also simulated an operational seamless design in which Stage~1 data are discarded from the confirmatory analysis. Under this design, dose selection follows the same two rules, but efficacy testing is a standard two-look group-sequential log-rank test on Stage~2 selected + control patients only, without combination testing or closed testing. The event targets under the operational design were 145 at IA and 216 at FA, chosen to match the expected Stage~2 event contributions in the inferential design under scenario A, so that both designs produce IA and FA at approximately the same calendar times.

For a direct comparison against a state-of-the-art alternative, we also implemented the multi-arm multi-stage design of \citet{zhang2025mams} on the same simulated data. The two procedures differ in how the Stage-1 (cohort-1) $p$-value is constructed for the intersection-hypothesis test: our procedure uses the cumulative cohort-1 log-rank statistic at each Phase 3 look and applies a single Hochberg correction, while \citet{zhang2025mams} construct a nested combination test in which the cohort-1 cumulative statistic at each Phase 3 look is decomposed into stage-wise incremental log-rank contributions, a Dunnett correction is applied to the incremental statistics at each stage separately (using the joint multivariate-normal distribution of the arm-vs-control statistics under the shared-control correlation), and the stage-specific Dunnett $p$-values are then combined across stages via an inverse-normal combination test. The Z\&J method therefore requires two sets of pre-specified weights, corresponding to the two levels of the nested combination: an outer set $h_{1j}, h_{2j}$ (eq.\ 2 of \citet{zhang2025mams}) combining cohort 1 with cohort 2 at each Phase 3 look $j$, which plays the same role as our $w_{k,1}, w_{k,2}$, and an inner set $w_{jl}$ (eq.\ 1 of \citet{zhang2025mams}) combining stage-wise incremental Dunnett $p$-values \emph{within} cohort 1 across Phase 3 looks $l = 2, \ldots, j$; the values used are given in Section~\ref{sim:results}. All other components (patient-wise separation, cohort-2 log-rank testing, group-sequential $\alpha$-spending, and dose selection rule) were kept identical between the two procedures.

Each combination of procedure, scenario, and selection rule was evaluated over $10{,}000$ independent replicates, with all three procedures applied to the same simulated data within each replicate, so that comparisons between procedures are paired.

\subsection{Base case: correctly specified weights}\label{sim:results}

In this subsection the combination weights of the two combination-test procedures are derived from the expected per-stage event share at each look under the primary planning scenario---scenario~A
---so that the design-time projections match the realized Stage-1 event counts at both looks; Section~\ref{sim:edge} then perturbs this specification.

For HB, we set $w_{k,1}^2 = E_{k,1}/(E_{k,1}+E_{k,2})$ for each look $k \in \{IA, FA\}$, with $E_{k,j}$ computed by integrating the accrual profile against scenario A's OS mixture.
This gave $E_{IA,1}=75.4$, $E_{IA,2}=144.6$, $E_{FA,1}=84.1$, $E_{FA,2}=215.9$ and correspondingly $w_{IA,1}=0.585$, $w_{IA,2}=0.811$, $w_{FA,1}=0.529$, $w_{FA,2}=0.848$.

For Z\&J, the outer weights were set equal to those of HB, $h_{1j} = w_{k,1}$ and $h_{2j} = w_{k,2}$, so that the two procedures share identical outer weights. At IA ($j=2$) the inner sum has a single term with $w_{2,2} = 1$, so the inner combination is trivial and the incremental Dunnett $p$-value coincides with a cumulative one. At FA ($j=3$) we set the two inner weights proportional to the square roots of the expected cohort-1 event contributions at each stage, giving $w_{3,2} = \sqrt{E_{IA,1}/E_{FA,1}}$ and $w_{3,3} = \sqrt{1 - E_{IA,1}/E_{FA,1}}$; using the same design-time projections $E_{IA,1}=75.4$ and $E_{FA,1}=84.1$ as HB, these evaluate to $w_{3,2} = 0.947$ and $w_{3,3} = 0.322$.

Table~\ref{tab:sim} reports the resulting rejection probabilities for the three procedures by scenario and selection rule.

\begin{table}[htbp]
\centering
\small
\caption{Rejection probability (\%) by testing procedure, scenario, and dose selection rule. \emph{Stage 2 only} is the operational seamless design, a conventional two-look group sequential log-rank test that discards Stage~1 data; \emph{HB with GS} is the proposed Hochberg-based closed testing procedure with group sequential boundaries; \emph{Z\&J} is the Dunnett-based procedure of \citet{zhang2025mams}. \emph{Selected} gives the probability that the high-dose arm was carried into Stage~2; the low-dose probability is its complement.\ 
IA = interim analysis rejection; FA = final analysis rejection (conditional on non-rejection at IA); Overall = IA + FA. All entries are percentages; Monte-Carlo standard errors are at most 0.50 percentage points (at most 0.16 on the null rows).
}
\label{tab:sim}
\begin{tabular}{ll r rrr rrr rrr}
\toprule
 & & \multicolumn{1}{c}{\textbf{Selected}} & \multicolumn{3}{c}{\textbf{Stage 2 only}} & \multicolumn{3}{c}{\textbf{HB with GS}} & \multicolumn{3}{c}{\textbf{Z\&J}} \\
\cmidrule(lr){3-3}\cmidrule(lr){4-6}\cmidrule(lr){7-9}\cmidrule(lr){10-12}
Scenario & Selection & High & IA & FA & Overall & IA & FA & Overall & IA & FA & Overall \\
\midrule
\multicolumn{12}{l}{\emph{Global null (target one-sided $\alpha=2.5\%$)}} \\
null & ORR & 50.40 & 0.64 & 1.95 & 2.59 & 0.86 & 1.48 & 2.34 & 0.86 & 1.50 & 2.36 \\
null & OS  & 50.08 & 0.64 & 1.98 & 2.62 & 0.89 & 1.47 & 2.36 & 0.88 & 1.49 & 2.37 \\
\midrule
\multicolumn{12}{l}{\emph{Equally-effective alternatives}} \\
A & ORR & 50.32 & 48.18 & 35.01 & 83.19 & 74.16 & 18.10 & 92.26 & 73.99 & 18.13 & 92.12 \\
A & OS  & 50.39 & 48.19 & 34.87 & 83.06 & 74.75 & 17.63 & 92.38 & 74.61 & 17.75 & 92.36 \\
B & ORR & 50.40 & 53.61 & 26.01 & 79.62 & 74.81 & 14.74 & 89.55 & 74.64 & 14.80 & 89.44 \\
B & OS  & 49.77 & 53.27 & 26.07 & 79.34 & 75.47 & 14.20 & 89.67 & 75.43 & 14.20 & 89.63 \\
C & ORR & 50.40 & 55.43 & 19.50 & 74.93 & 72.23 & 11.09 & 83.32 & 71.93 & 11.11 & 83.04 \\
C & OS  & 49.88 & 54.59 & 19.88 & 74.47 & 73.70 & 10.62 & 84.32 & 73.64 & 10.56 & 84.20 \\
\midrule
\multicolumn{12}{l}{\emph{Differential alternatives (low dose $<$ high dose)}} \\
A1 & ORR & 99.09 & 47.83 & 34.31 & 82.14 & 64.84 & 22.31 & 87.15 & 66.90 & 21.67 & 88.57 \\
A1 & OS  & 86.13 & 41.42 & 30.14 & 71.56 & 57.88 & 18.91 & 76.79 & 59.82 & 18.34 & 78.16 \\
A2 & ORR & 89.00 & 43.88 & 33.19 & 77.07 & 63.27 & 21.61 & 84.88 & 64.28 & 20.88 & 85.16 \\
A2 & OS  & 72.13 & 37.24 & 30.40 & 67.64 & 55.76 & 21.27 & 77.03 & 56.62 & 20.38 & 77.00 \\
A3 & ORR & 73.22 & 41.08 & 35.04 & 76.12 & 63.94 & 22.46 & 86.40 & 64.27 & 22.31 & 86.58 \\
A3 & OS  & 62.13 & 38.25 & 34.39 & 72.64 & 61.54 & 22.84 & 84.38 & 61.88 & 22.63 & 84.51 \\
\bottomrule
\end{tabular}
\end{table}

\paragraph{FWER control.} Under the global null, all three procedures controlled the one-sided FWER at the nominal $\alpha=2.5\%$, with estimates of 2.34\%--2.62\% across procedures and selection rules (Monte-Carlo standard error $\leq 0.16$ percentage points). Both HB with GS and Z\&J methods demonstrated an FWER lower than 2.5\%. Because the boundary rule of Section~\ref{sim:design} spends exactly $\alpha=0.025$ across the two looks in every replicate, the residual conservatism of the two combination-test procedures is a property of the closed test and not of the boundary calculation. 

\paragraph{HB with GS vs.\ Z\&J vs.\ Stage 2 only.} Taking the three procedures together, the operational seamless design was uniformly the weakest, while the two combination-test procedures were close enough to each other.


Although HB and Z\&J gave nearly identical operating characteristics throughout, we note that
their difference in overall power is ordered by how dissimilar the two experimental arms are. Under the equally-effective scenarios A, B and C, where the arms are exchangeable, Z\&J was marginally \emph{behind} HB, by $0.28$ percentage points or less in every cell. Under the differential scenarios the ordering reverses, but the reversal is material only once the arms are far apart: the gap does not exceed $+0.28$ percentage points in A2 (low-dose RM $=0.80$) or A3 (RM $=0.70$), in both of which it is within Monte-Carlo error of zero, and rises to about $+1.4$ percentage points in A1 (low dose null) under both selection rules ($+1.42$ for ORR-based and $+1.37$ for OS-based selection, paired Monte-Carlo SE $\approx 0.15$). Scenario A1 is thus the only cell in which the two procedures are distinguishable beyond Monte-Carlo error.

This gradient is what the two multiplicity corrections predict. A Dunnett correction exploits the joint multivariate-normal distribution of the arm-vs-control statistics under the shared-control correlation, and its advantage over a Hochberg step-up correction is greatest when the arm-vs-control statistics are most dispersed, which is precisely the configuration of A1. When the arms are exchangeable there is no dispersion to exploit and the extra machinery returns nothing. The gain is also concentrated at the interim: in A1 the Z\&J advantage at IA is about $+2$ percentage points under both rules ($+2.06$ for ORR-based and $+1.94$ for OS-based selection), of which about a third is given back at FA ($-0.64$ and $-0.57$), because the incremental-Dunnett construction places weight on a Stage-1 increment that contributes few additional events between the two looks. This is the same mechanism that makes Z\&J the more fragile of the two under weight misspecification, as Section~\ref{sim:edge} shows.

Elsewhere the two are interchangeable in operating characteristics, so the choice between them can be made on grounds other than statistical efficiency. The HB approach\ 
needs only one set of combination weights rather than the two nested sets Z\&J requires --- a practical consideration given that the weights must be committed at design time and, as Section~\ref{sim:edge} shows, are the component most exposed to misprojection.

\paragraph{ORR-based vs.\ OS-based selection.} Under the differential scenarios A1--A3, the ORR-based selection rule outperformed the OS-based rule for both combination-test procedures, with a higher probability of selecting the high-dose arm and greater power. 
The power advantage narrowed monotonically as the two experimental arms became more similar.
\ This means that in such a response-mediated setting the ORR comparison---where 75\% of the accrued Stage-1 patients contribute a response observation---is better able to identify the more effective dose than the immature survival comparison available at the same data cut-off (an expected 27\%--31\% across scenarios A1--A3 contribute a death).

Under the equally-effective alternative scenarios A, B and C, the two rules give almost the same dose selection probabilities. 
This is expected: when both experimental arms have identical true distributions, the choice of selection rule cannot systematically favour one dose over the other. Scenario C is a marginal exception for power, with OS-based selection achieving higher power by about $1$ percentage point for both HB with GS and Z\&J methods (1.00 and 1.16 percentage points respectively, paired Monte-Carlo SE 0.27). This means that although the two arms have identical marginal distributions, what a selection rule can still do for power is to favour the arm whose \emph{realized} Stage-1 survival --- not yet fully observed at the time of selection --- happens to be the better. The behaviour of the ORR-based rule depends on the correlation between ORR and OS, and as scenario C lengthens the non-responder median from 4.04 to 7.53~months against an unchanged 22.81~months for responders, that Spearman correlation falls from 0.60 and 0.55 in scenarios A and B to 0.40. By contrast, the OS-based rule reads survival at the time of selection directly and is affected less. The gap between the two rules is therefore most visible in scenario C.

\subsection{Edge case: misspecified weights}\label{sim:edge}

The cohort combination weights $(w_{k,1}, w_{k,2})$ used by our procedure, the corresponding cohort weights $(h_{1j}, h_{2j})$ and the inner cross-stage weights $(w_{jl})$ used by Z\&J, and the IA group-sequential boundary $Z_{IA}^*$ are all fixed at design time from the \emph{expected} per-stage event counts; only $Z_{FA}^*$ is re-solved at analysis time, and then at the \emph{realized} correlation (Section~\ref{sim:design}). In event-driven trials with time-to-event endpoints, however, actual event accrual can diverge from the design projection, particularly in disease settings with long-tail survival. The projection that is hardest to get right is not the Stage-1 event total at the FA---Stage-1 patients enrol first, so their follow-up is the most mature of any group in the trial---but the point along that accrual at which the interim falls, and hence how the Stage-1 total divides into the portion already available at the IA and the increment arriving between the two looks. 
To assess how each procedure behaves when that division is misprojected, we repeated the head-to-head comparison of Section~\ref{sim:results} under a misspecified design in which the analyst \emph{believes} that Stage-1 patients will have contributed $E_{IA,1}=64.1$ events by the IA out of a correctly projected $E_{FA,1}=84.1$ at FA---a 20-event increment between the two looks---whereas the underlying data-generating process delivers about 75 events by the IA and about 84 by the FA, an increment of only about 9. 
The Stage-1 total is therefore projected correctly and only its timing relative to the interim is wrong. Under this misspecification HB updates its IA cohort weight to $w_{IA,1}=0.540$ (from $0.585$ under correct specification) and leaves $w_{FA,1}=0.529$ untouched, while Z\&J additionally updates its FA inner cross-stage weights to $(w_{3,2}, w_{3,3}) = (0.873, 0.488)$ (from $(0.947, 0.322)$), placing about 24\% of the FA cohort-1 weight on the incremental Stage-1 piece (up from about 10\%). The IA boundary, by contrast, is untouched: the identity $R^{2}=D_{IA}/D_{FA}$ holds whenever the cohort weights are set from the same projected event counts that enter the information fraction, however wrong those counts are, so the misspecified design locks in the same IA boundary $Z_{IA}^{*}=2.371$ as the correctly-specified one, for HB and Z\&J alike. Both procedures use the misspecified expected event counts to fix weights and boundaries but observe the correctly-generated events under the data-generating scenarios of Section~\ref{sim:design} at analysis time.

Table~\ref{tab:hb-vs-zhang-edge} reports the overall rejection probabilities under misspecification. Both procedures preserved the one-sided type 1 error rate at the nominal $\alpha=2.5\%$ under the null (2.36\% for HB and 2.31\% for Z\&J), confirming that the inverse-normal combination is robust to a mismatch between the assumed and realized weights on the null. Under the alternatives the two procedures separate cleanly. HB was essentially unaffected: compared replicate by replicate against its correctly-specified counterpart it moved by at most $0.28$ percentage points in either direction, and the only two changes exceeding Monte-Carlo error were small \emph{gains}, in scenario C ($+0.28$, paired SE $0.07$) and scenario A1 ($+0.10$, paired SE $0.04$). Z\&J lost power in all six alternatives, by between $0.57$ and $1.03$ percentage points against paired standard errors of $0.10$--$0.14$, so each of those losses is real. The mechanism is that Z\&J must commit at design time a weight for the Stage-1 \emph{increment} between the two looks, and that increment is a small difference of two much larger event totals: a 15\% error in the projected Stage-1 count at the IA (64.1 against a realized 75) turns a projected 20-event increment into a realized 9, an error of 122\%. The combination weights are then no longer proportional to the information the two Stage-1 pieces actually carry. 
HB never has to project that split, because it uses the cumulative Stage-1 log-rank statistic at each look and its FA cohort weight depends only on the Stage-1 event \emph{total}, which this misspecification gets right.

\begin{table}[htbp]
\centering
\caption{Head-to-head comparison of the proposed Hochberg-based procedure (HB) against the Dunnett-based procedure of \citet{zhang2025mams} (Z\&J), under a misspecified design in which weights and boundaries are fixed from an event projection that gets the Stage-1 FA event total right ($E_{FA,1}=84.1$) but places too few of those events before the interim ($E_{IA,1}=64.1$ vs.\ realized $\approx 75$), and so assumes a 20-event Stage-1 increment between the two looks where only about 9 materialise. Values are overall rejection probabilities (\%) with ORR-based dose selection. Baseline columns reproduce the correctly-specified results of Table~\ref{tab:sim} for reference; Monte-Carlo standard errors are at most 0.43 percentage points (at most 0.16 on the null row).}
\label{tab:hb-vs-zhang-edge}
\begin{tabular}{lrrrr}
\toprule
 & \multicolumn{2}{c}{\textbf{Baseline (correct spec)}} & \multicolumn{2}{c}{\textbf{Edge (misspecified spec)}} \\
\cmidrule(lr){2-3}\cmidrule(lr){4-5}
Scenario & HB & Z\&J & HB & Z\&J \\
\midrule
null  &  2.34 &  2.36 &  2.36 &  2.31 \\
\midrule
A     & 92.26 & 92.12 & 92.26 & 91.55 \\
B     & 89.55 & 89.44 & 89.57 & 88.67 \\
C     & 83.32 & 83.04 & 83.60 & 82.41 \\
\midrule
A1    & 87.15 & 88.57 & 87.25 & 87.81 \\
A2    & 84.88 & 85.16 & 84.91 & 84.43 \\
A3    & 86.40 & 86.58 & 86.39 & 85.55 \\
\bottomrule
\end{tabular}
\end{table}

These results reveal a robustness contrast that the correctly-specified comparison in Section~\ref{sim:results} did not surface. Under correct specification the two procedures were essentially interchangeable, with Z\&J marginally ahead in the one scenario, A1, where the arm-vs-control statistics are dispersed enough for the Dunnett correction to pay. Misprojecting only the \emph{timing} of the Stage-1 events---not their total---
leaves HB where it was and costs Z\&J something on every alternative. 
On the equally-effective alternatives A, B and C, where HB led by at most $0.28$ percentage points, its lead widens to $0.71$, $0.90$ and $1.19$. On A2 and A3 the ordering reverses outright, margins of $0.28$ and $0.18$ percentage points in Z\&J's favour becoming deficits of $0.48$ and $0.84$. In A1---the one cell in which the Dunnett correction genuinely pays---the ordering survives, but the margin falls from $+1.42$ to $+0.56$ percentage points, that is from nine paired standard errors to three. 
Z\&J is 
more fragile when the assumed increment fails to materialise on schedule. Practitioners choosing between these procedures should weigh both statistical efficiency and robustness to event-projection uncertainty, particularly in disease settings with long-tail survival where the position of the interim within the Stage-1 event accrual is difficult to project accurately.

\subsection{Sufficiency of individual-hypothesis boundaries for the intersection test}\label{sim:alphaint}

Section~\ref{sec3} describes an iterative $\gamma$-calibration step that tightens the FA boundary of the intersection-hypothesis test whenever the value of $\alpha(Z_{IA}^{*},Z_{FA}^{*})$ exceeds the nominal $\alpha$. Whether this step is actually needed depends on the combination weights and on the information fractions realized at analysis time, neither of which is known exactly at design time. Here we evaluate $\alpha(Z_{IA}^{*},Z_{FA}^{*})$ using the deterministic quadrature (\ref{eq18}) of Appendix~\ref{app:alphaint} over the configurations arising in the simulation study.

Writing $\mathrm{IF}_j = (e_{IA,j,i^{*}}+e_{IA,j,0})/(e_{FA,j,i^{*}}+e_{FA,j,0})$ for the stage-$j$ information fraction, the correlation (\ref{eq:corr}) is $R=\mathrm{Corr}(c_{IA},c_{FA}) = w_{IA,1}w_{FA,1}\sqrt{\mathrm{IF}_1}+w_{IA,2}w_{FA,2}\sqrt{\mathrm{IF}_2}$. 
In our event-driven design the event targets are fixed at $D_{IA}=220$ and $D_{FA}=300$, and hence the planned information fraction at $D_{IA}/D_{FA}=0.733$. 
However, the two stage-wise fractions still vary substantially and in opposite directions, and so therefore does the realized $R$. 

Because Stage-1 patients enrol first, their follow-up is the most mature and their FA event contribution is the best-projected quantity in the design, so we assume $E_{FA,1}=84.1$ is held fixed while $\mathrm{IF}_1$ is varied; with the event targets $D_{IA}$ and $D_{FA}$ also fixed this determines $\mathrm{IF}_2$ and hence $R$, so $R$ is a function of $\mathrm{IF}_1$ alone. The sweep therefore varies $\mathrm{IF}_1$. 
The weights, by contrast, are not determined by the data: they are fixed at design time and enter $\alpha$ directly as well as through $R$. 
We explore five weights configurations: the planned design of Section~\ref{sim:results}, the misspecified of Section~\ref{sim:edge}, and three probe values spanning stage-1-heavy through stage-2-heavy choices. 

\begin{figure}[htbp]
\centering
\includegraphics[width=0.78\textwidth]{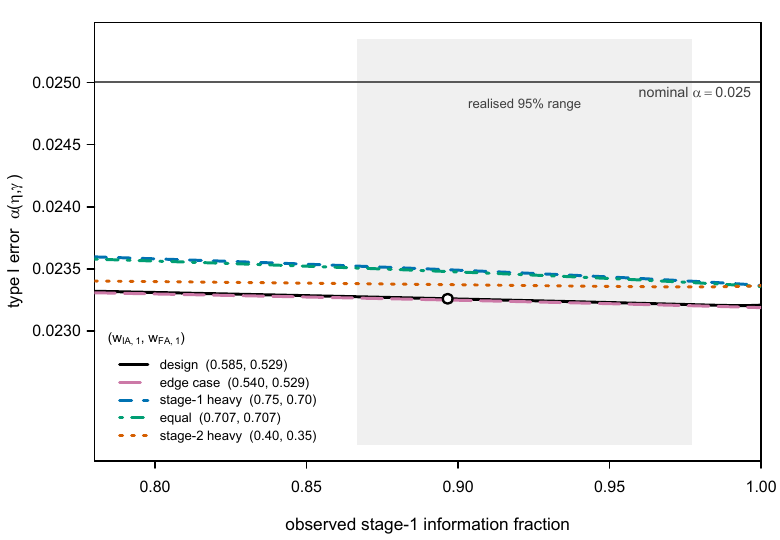}
\caption{Type~1 error rate of the repeated intersection-hypothesis test, $\alpha(\eta,\gamma)$, when the boundaries are set equal to those for the individual hypothesis, $\eta=Z_{IA}^{*}$ and $\gamma=Z_{FA}^{*}$, evaluated by (\ref{eq18}) at $\rho=0.5$. 
The shaded band gives the central 95\% range of $\mathrm{IF}_1$ over the global-null replicates of Section~\ref{Simulation Studies}, and the horizontal line is the nominal level $\alpha=0.025$.}
\label{fig:alphaweights}
\end{figure}

Figure~\ref{fig:alphaweights} shows the resulting type~1 error rate for the intersection hypothesis against different realized $\mathrm{IF}_1$ under different weight specifications. Across all five curves $\alpha(Z_{IA}^{*},Z_{FA}^{*})$ lies between $0.0232$ and $0.0236$, leaving at least $1.4\times 10^{-3}$ of the nominal $\alpha=0.025$ unspent, and the shaded band indicates that the configurations actually realized in the simulation occupy only a small part of that region: over the realized range of $\mathrm{IF}_1$ the five curves span $0.02321$--$0.02352$. 
Rejecting the intersection hypothesis using the boundaries calculated for the individual hypothesis therefore preserves the type~1 error rate at level $\alpha$ throughout, with no further calibration of $\gamma$ required. 
We emphasize that this is a numerical finding for the design class considered and not a general guarantee. 

\section{Summary and Discussion}\label{summary}

In this paper we proposed a testing procedure for adaptive seamless Phase~II/III oncology trials with a time-to-event primary endpoint, in which dose selection is followed by group sequential repeated testing during Phase~III. The procedure combines patient-wise separation for combination $p$-values \citep{jenkins2011adaptive,di2011time,carreras2015adaptive} with a closed testing procedure, and derives dedicated group sequential boundaries for the individual hypothesis and the intersection hypothesis separately. For the individual hypothesis, the joint distribution of the IA and FA inverse-normal combination statistics is bivariate normal under the null, so boundaries can be obtained from standard group sequential software. For the intersection hypothesis, the joint distribution is not multivariate normal, and we derived a numerical integration approach in Section~\ref{sec3} that exploits symmetry of the underlying multivariate normal to reduce the type~1 error to a linear combination of quadrivariate normal probabilities. This is where the present construction departs from existing proposals for repeated testing after an adaptation interim, which index the boundaries for the post-adaptation looks by the overall information fraction of the primary endpoint. That is the correct index for the canonical joint distribution, in which the successive statistics are those of an unadapted trial; it is not the correct index for the combination statistics that are actually tested, and for the intersection hypothesis there is no correlation available to index at all, the joint distribution not being multivariate normal. Evaluating the type~1 error directly from that joint distribution dispenses with the index, and with it the need to support FWER control by simulation alone. Lemma~\ref{lem:key} justifies calibrating the boundaries independently for each hypothesis and then applying closed testing at each look while retaining FWER control.

The simulation study illustrates two points about adaptive dose selection. First, when both experimental arms are truly equally effective (scenarios A, B, C), the two rules give almost identical dose selection probabilities, since the arms are exchangeable and no rule can systematically identify a ``better'' arm; power follows suit in scenarios A and B, where the two rules are within Monte-Carlo error of one another, while in scenario C the OS-based rule retains a small advantage of about one percentage point by favouring the arm whose \emph{realized} Stage-1 survival happens to be the better. Second, when the low-dose and high-dose arms differ in true efficacy (scenarios A1--A3), the ORR-based rule yields modest but consistent power gains over the OS-based rule, and the magnitude of the gain scales with the true effect gradient between the two doses. In the extreme case where only the high-dose arm is effective (scenario A1), an ORR-based rule that correctly identifies the effective arm on 99\% of replicates delivered 87\% overall power, compared to 77\% for the OS-based rule. The comparison to an operational seamless design that discards Stage~1 data highlights a complementary point: retaining Stage~1 patients through combination testing recovers approximately 8--10 percentage points of power under equally-effective alternatives, and the value of doing so is largest when the selection rule is imperfect (e.g. OS-based selection under differential alternatives).

Head-to-head against the Dunnett-based procedure of \citet{zhang2025mams}, the two constructions gave nearly identical operating characteristics under correct design specification, so the choice between them turns on robustness rather than on statistical efficiency at the planned configuration. Under a design that projects the Stage-1 event total correctly but misplaces its division either side of the interim (Section~\ref{sim:edge}), the incremental-Dunnett construction of \citet{zhang2025mams} loses 0.6--1.0 percentage points of power on every alternative considered, whereas the proposed Hochberg construction remains robust; it weights only cumulative Stage-1 totals and never has to project a split. Nor does that advantage depend on the design staying at two looks. Because its covariance is defined increment by increment, the construction of \citet{zhang2025mams} carries over to more analysis times without further derivation, and might on that ground seem the more extensible of the two; but the closed-form simplification of Section~\ref{sec3} is not the only route to the intersection-hypothesis boundaries, and the recursion of Appendix~\ref{app:Llook} evaluates the same type 1 error at any number of looks at a cost linear in that number. Extensibility to further looks therefore does not separate the two constructions, whereas robustness continues to favour ours: the incremental construction carries one Stage-1 piece per look rather than the two of the present design, and each further cut is a smaller difference of two much larger cumulative event counts, so the same absolute error in the projected event trajectory becomes a proportionally larger error in the weight that piece must be given, and the fragility documented in Section~\ref{sim:edge} would be expected to grow with the number of looks.

Several limitations warrant mention. First, the methodology is developed for the case of two experimental doses. The general procedure in Lemma~\ref{lem:key} extends to $F\geq 2$ hypotheses and $J\geq 2$ analysis times, and Appendix~\ref{app:Llook} lifts the two-look restriction of the closed-form simplification in Section~\ref{sec3}, but both that simplification and the recursion exploit the two-arm structure---a single shared-control correlation, which one fixed rotation diagonalizes---so extension to three or more experimental doses will require corresponding derivations. That restriction is less consequential in practice than it may appear. Phase~III oncology trials are large and expensive, which keeps the number of experimental doses carried into the confirmatory stage small; two is the usual number in a seamless design following dose optimization, and three or more are rare.
Second, comparison with other multiplicity control strategies --- notably the graphical approach of \citet{sugitani2016simple}, the conditional error function approach of \citet{koenig2008adaptive}, and Bonferroni-style $\alpha$-splitting across looks --- was outside the scope of the present simulation and is a natural next step.

Overall, the proposed method offers a principled path to combine seamless Phase~II/III dose selection with group sequential repeated testing for time-to-event endpoints, an area in which existing methodology has largely been confined to two-stage designs without repeated Phase~III testing. The framework supports designs that can accelerate confirmation of efficacy after dose optimization, consistent with the goals of Project Optimus.

\appendix

\section{A FWER-preserving procedure for boundaries nested within CTP}\label{app:lemma1}

This appendix states and proves the general result underlying
Section~\ref{sec:general}: nesting group-sequential boundaries within the
closed testing procedure controls the FWER in the strong sense, for any
number of hypotheses and any number of analysis times. We keep the sign
convention of Section~\ref{sec:gsd} throughout: $p$-values enter the test
statistics through $\Phi^{-1}$, so \emph{small} values are evidence
against the null and a hypothesis is rejected when its statistic falls at
or below the corresponding boundary, as in (\ref{eq1})--(\ref{eq2}) and
(\ref{eq9}). The hypotheses $H_1,\dots,H_F$ are nulls, one for each
comparison of interest; in the design of Section~\ref{sec:setup}, $F=2$ and
$H_f$ is $H_{0,f}:\theta_f=1$, the hypothesis of no effect of dose $f$
versus control. Writing
$\mathcal{T}_0=\{f:\ H_f\ \text{is true}\}$
for the (unknown) index set of the true hypotheses, the familywise error rate is
the probability of rejecting at least one of them,
\[
    \mathrm{FWER}\ =\ Pr\!\left(\bigcup_{f\in\mathcal{T}_0}\bigl\{H_f\ \text{is rejected}\bigr\}\right),
\]
and control at level $\alpha$ is \emph{strong} if $\mathrm{FWER}\leq\alpha$
for every configuration $\mathcal{T}_0$, as against \emph{weak} control,
which requires the bound only under the global null
$\mathcal{T}_0=\{1,\dots,F\}$.

\begin{lemma}\label{lem:key}
Assume one wants to perform one-sided tests against multiple hypotheses
$H_1,\dots,H_F$ ($F\geq 2$) at analysis times $j=1,\dots,J$ ($J\geq 2$). If
\begin{enumerate}
    \item for each intersection hypothesis $H_S \triangleeq \bigcap_{f \in S} H_f$ with $\emptyset\neq S\subseteq\{1,\dots,F\}$ and each analysis time $j$, a test statistic $T_{j,S}$ against $H_S$ is available, and
    \item for each $H_S$ there is a sequence of boundaries $(C_{j,S})_{j=1}^J$ whose repeated use is of level $\alpha$ for $H_S$, that is
    \[
        \sup_{P\in H_S}\ Pr_P\!\left(\exists\, j\in\{1,\dots,J\}:\ T_{j,S}\leq C_{j,S}\right)\ \leq\ \alpha,
    \]
    equivalently $\inf_{P\in H_S} Pr_P(T_{1,S}>C_{1,S},\dots,T_{J,S}>C_{J,S})\geq 1-\alpha$ in the form of (\ref{eq9}), the supremum running over all distributions $P$ compatible with the composite null $H_S$,
\end{enumerate}
then the following procedure controls the FWER at level $\alpha$ in the
strong sense:
\begin{enumerate}
    \item Start at $j=1$.
    \item For every $f\in\{1,\dots,F\}$ not already rejected: if for each $S$ with $f\in S$ there exists a look $k_S\leq j$ such that $T_{k_S,S}\leq C_{k_S,S}$, then reject $H_f$.
    \item Set $j\leftarrow j+1$; if $j>J$ stop, else return to step~2.
\end{enumerate}
Rejections are retained once made, and the crossings that license a
rejection of $H_f$ may occur at different looks $k_S$ for different $S$.
\end{lemma}

The trial may of course be terminated once a hypothesis of interest has
been rejected; the bound below holds whatever such stopping rule is
adopted, since stopping can only remove rejection opportunities.

\begin{proof}
Assume $\mathcal{T}_0\neq\emptyset$, since otherwise no true hypothesis
exists and the FWER is zero. Being a non-empty subset of
$\{1,\dots,F\}$, $\mathcal{T}_0$ indexes one of the intersection
hypotheses of condition~1, namely
$H_{\mathcal{T}_0}=\bigcap_{f\in\mathcal{T}_0}H_f$.

Fix any $f^{*}\in\mathcal{T}_0$ and any look $j$. By step~2, the procedure
rejects $H_{f^{*}}$ at look $j$ only if, for every $S$ containing $f^{*}$,
there is a look $k_S\leq j$ with $T_{k_S,S}\leq C_{k_S,S}$. Since
$f^{*}\in\mathcal{T}_0$, the set $\mathcal{T}_0$ is one such $S$, whence
\[
\bigl\{H_{f^{*}}\ \text{is rejected at look}\ j\bigr\}
\ \subseteq\ A_j\ \triangleeq\ \bigl\{\exists\, k\leq j:\ T_{k,\mathcal{T}_0}\leq C_{k,\mathcal{T}_0}\bigr\}.
\]
The events $A_1\subseteq A_2\subseteq\dots\subseteq A_J$ do not depend on
$f^{*}$: the single intersection hypothesis $H_{\mathcal{T}_0}$ serves every
true hypothesis at once, precisely because $\mathcal{T}_0$ contains all of
them. A hypothesis is rejected only if it is rejected at some look, so
taking the union over $f^{*}\in\mathcal{T}_0$ and over $j\in\{1,\dots,J\}$
and using the nesting,
\[
\bigcup_{f^{*}\in\mathcal{T}_0}\bigl\{H_{f^{*}}\ \text{is rejected}\bigr\}
\ =\ \bigcup_{f^{*}\in\mathcal{T}_0}\ \bigcup_{j=1}^{J}\bigl\{H_{f^{*}}\ \text{is rejected at look}\ j\bigr\}
\ \subseteq\ \bigcup_{j=1}^{J}A_j\ =\ A_J,
\]
and condition~2, applied to the true hypothesis $H_{\mathcal{T}_0}$, bounds
$Pr(A_J)$ by $\alpha$. Hence
\[
\mathrm{FWER}\ =\ Pr\!\left(\bigcup_{f^{*}\in\mathcal{T}_0}\bigl\{H_{f^{*}}\ \text{is rejected}\bigr\}\right)
\ \leq\ Pr(A_J)\ =\ Pr\!\left(\exists\, j\in\{1,\dots,J\}:\ T_{j,\mathcal{T}_0}\leq C_{j,\mathcal{T}_0}\right)\ \leq\ \alpha.
\]
No property of $\mathcal{T}_0$ was used beyond its being the index set of
the true hypotheses, so the bound holds for every configuration of true and false
hypotheses and control is strong.
\end{proof}

The
argument above is that of \citet{marcus1976closed}, adapted to repeated
testing: the local test of $H_S$ is ``reject if $T_{j,S}$ ever falls at or
below $C_{j,S}$'', and condition~2 is exactly the statement that this
local test is of level $\alpha$.

\begin{remark}
Lemma~\ref{lem:key} reduces FWER control to a finite set of level-$\alpha$
verifications, one for each intersection hypothesis: although the proof
invokes condition~2 only at $S=\mathcal{T}_0$, that set is unknown at the
design stage, so the condition has to be verified for every $S$. In the
two-dose, two-look design of Section~\ref{method} there are two cases.
Under the global null,
$\mathcal{T}_0=\{1,2\}$ and the relevant $S$ is the full intersection,
whose repeated test is built from (\ref{eq7})--(\ref{eq8}); condition~2
is then the requirement $\alpha(Z^{*}_{IA},Z^{*}_{FA})\leq\alpha$ of
Section~\ref{sec3}, computed in Appendix~\ref{app:alphaint}. If exactly
one dose is truly ineffective,
$\mathcal{T}_0$ is a singleton $\{f\}$ and the relevant $S$ is that
singleton, whose repeated test is built from (\ref{eq5})--(\ref{eq6}).
For the singleton the verification needs care, because the hypothesis index
$f$ is fixed whereas the arm carried into Stage~2 is not. Take the singleton
statistic to be built from the Stage-1 $p$-value of the \emph{fixed} arm $f$,
that is $p_{k,1,f}$, rather than that of the selected arm $p_{k,1,i^{*}}$.
With this choice $\Phi^{-1}(p_{k,1,f})$ is
standard normal, and conditionally on the Stage-1 data---hence on the
selection, which is a function of it---patient-wise separation leaves
$\Phi^{-1}(p_{k,2})$ standard normal and independent of it, so the IA and FA
statistics are bivariate normal with the covariance (\ref{eq:corr}) and their
repeated use at $(Z^{*}_{IA},Z^{*}_{FA})$ is of level $\alpha$. The procedure
of Section~\ref{sec:setup} tests only the selected arm, and does so with
$p_{k,1,i^{*}}$; the two statistics coincide on $\{i^{*}=f\}$, while on the
complement arm $f$ has been dropped, no Stage-2 data accrue on it, and we
set $p_{k,2}=1$ by convention, so that $H_{0,f}$ can never be rejected.
The local test
of $H_{0,f}$ that the procedure performs therefore rejects only on the
intersection of $\{i^{*}=f\}$ with a crossing of the fixed-arm statistic, a
subset of the latter event, and condition~2 for singletons follows;
Lemma~\ref{lem:key} then bounds the FWER by $\alpha$ in this configuration.
\end{remark}

\section{Numerical computation of \texorpdfstring{$\alpha(\eta,\gamma)$}{alpha(eta, gamma)}}\label{app:alphaint}

This appendix derives the deterministic expression for the intersection-hypothesis type~1 error function $\alpha(\eta,\gamma)$ referenced in Section~\ref{sec3}. The setup is that of Section~\ref{sec:inter}, with intersection combination statistics (\ref{eq7}) and (\ref{eq8}) evaluated at candidate boundaries $\eta$ (IA) and $\gamma$ (FA).

\subsection{Conditioning on Stage-2 statistics}

Stage 1 and Stage 2 data are independent by patient-wise separation, and the two Stage-2 statistics follow a bivariate normal distribution with mean zero, unit variance and covariance $\sqrt{(e_{IA,2,i^{*}}+e_{IA,2,0})/(e_{FA,2,i^{*}}+e_{FA,2,0})}$, i.e.\ the square root of the Stage-2 information fraction at IA. Denoting this density by $f(\zeta_{IA},\zeta_{FA})$, the left-hand side of (\ref{eq9}) becomes
\begin{align}
& Pr\left(c_{IA}(p_{IA,1}^{HB},p_{IA,2})>\eta,\ c_{FA}(p_{FA,1}^{HB},p_{FA,2})>\gamma\right) \nonumber\\
& = \int\!\!\int_{\zeta_{IA},\zeta_{FA}} Pr\!\left(\Phi^{-1}(p_{IA,1}^{HB})>\tfrac{\eta-w_{IA,2}\zeta_{IA}}{w_{IA,1}},\ \Phi^{-1}(p_{FA,1}^{HB})>\tfrac{\gamma-w_{FA,2}\zeta_{FA}}{w_{FA,1}}\right) f(\zeta_{IA},\zeta_{FA})\,d\zeta_{IA}\,d\zeta_{FA}. \label{eq10}
\end{align}

Write $t=(\eta-w_{IA,2}\zeta_{IA})/w_{IA,1}$, $s=(\gamma-w_{FA,2}\zeta_{FA})/w_{FA,1}$, and let $Z_{IA,(1)}=\min(Z_{IA,1,1},Z_{IA,1,2})$, $Z_{IA,(2)}=\max(Z_{IA,1,1},Z_{IA,1,2})$ (analogously $Z_{FA,(1)},Z_{FA,(2)}$). Because $p_{k,1}^{HB}$ is the Hochberg-adjusted $p$-value of the two arm-vs-control statistics, the integrand of the right of (\ref{eq10}) excluding $f(\zeta_{IA},\zeta_{FA})$ can be written as
\begin{align}
& Pr\!\left(\Phi^{-1}(p_{IA,1}^{HB})>t,\ \Phi^{-1}(p_{FA,1}^{HB})>s\right) \nonumber\\
& =\ Pr\!\left(Z_{IA,(1)}>\Phi^{-1}(\Phi(t)/2),\ Z_{IA,(2)}>t,\ Z_{FA,(1)}>\Phi^{-1}(\Phi(s)/2),\ Z_{FA,(2)}>s\right) \label{eq11}\\
& =\ Pr\!\left(Z_{IA,(1)}<-t,\ Z_{IA,(2)}<\Phi^{-1}(1-\Phi(t)/2),\ Z_{FA,(1)}<-s,\ Z_{FA,(2)}<\Phi^{-1}(1-\Phi(s)/2)\right), \label{eq12}
\end{align}
where (\ref{eq11}) follows from the definition of $p^{HB}$ and (\ref{eq12}) uses the symmetry of the joint distribution of $(Z_{IA,1,1},Z_{IA,1,2},Z_{FA,1,1},Z_{FA,1,2})$ under the null:
\begin{align} \label{dist1}
\begin{pmatrix}Z_{IA,1,1}\\ Z_{IA,1,2}\\ Z_{FA,1,1}\\ Z_{FA,1,2}\end{pmatrix} \sim N\!\left(\mathbf{0},\ \begin{pmatrix}1 & \rho & r & \rho r\\ \rho & 1 & \rho r & r\\ r & \rho r & 1 & \rho\\ \rho r & r & \rho & 1\end{pmatrix}\right),
\end{align}
with $\rho=0.5$ the arm-vs-arm correlation due to the shared control arm \citep{dunnett1955multiple} and $r=\sqrt{(e_{IA,1,i^{*}}+e_{IA,1,0})/(e_{FA,1,i^{*}}+e_{FA,1,0})}$ the within-arm cross-look correlation.

\subsection{Reduction to a five-term quadrivariate probability}

Let $u_1=-t$, $u_2=\Phi^{-1}(1-\Phi(t)/2)$, $v_1=-s$, $v_2=\Phi^{-1}(1-\Phi(s)/2)$, and $\xi_{IA,i}=Z_{IA,1,i}$, $\xi_{FA,i}=Z_{FA,1,i}$. Expanding the min/max events in (\ref{eq12}) via inclusion--exclusion and repeatedly invoking the arm-exchange symmetry of (\ref{dist1}) reduces the RHS to
\begin{align}
& Pr\!\left(Z_{IA,(1)}<u_1,Z_{IA,(2)}<u_2,Z_{FA,(1)}<v_1,Z_{FA,(2)}<v_2\right) \nonumber\\
& =\ 2Pr(\xi_{IA,1}<u_1,\xi_{IA,2}<u_2,\xi_{FA,1}<v_1,\xi_{FA,2}<v_2) + 2Pr(\xi_{IA,1}<u_1,\xi_{IA,2}<u_2,\xi_{FA,1}<v_2,\xi_{FA,2}<v_1) \nonumber\\
& \quad -\ 2Pr(\xi_{IA,1}<u_1,\xi_{IA,2}<u_2,\xi_{FA,1}<v_1,\xi_{FA,2}<v_1) - 2Pr(\xi_{IA,1}<u_1,\xi_{IA,2}<u_1,\xi_{FA,1}<v_1,\xi_{FA,2}<v_2) \nonumber\\
& \quad +\ Pr(\xi_{IA,1}<u_1,\xi_{IA,2}<u_1,\xi_{FA,1}<v_1,\xi_{FA,2}<v_1). \label{eq14}
\end{align}
Denote $F(a,b,c,d)=Pr(\xi_{IA,1}<a,\xi_{IA,2}<b,\xi_{FA,1}<c,\xi_{FA,2}<d)$ the quadrivariate normal CDF of (\ref{dist1}). Substituting (\ref{eq14}) back into (\ref{eq10}) gives the type~1 error
\begin{align}
\alpha(\eta,\gamma) = 1 - \int\!\!\int_{\zeta_{IA},\zeta_{FA}} \big\{ & 2F(u_1,u_2,v_1,v_2) + 2F(u_1,u_2,v_2,v_1) - 2F(u_1,u_2,v_1,v_1) \nonumber\\
& - 2F(u_1,u_1,v_1,v_2) + F(u_1,u_1,v_1,v_1) \big\} f(\zeta_{IA},\zeta_{FA})\,d\zeta_{IA}\,d\zeta_{FA}. \label{eq15}
\end{align}
Each of the five terms in (\ref{eq15}) is a quadrivariate normal probability. In principle these must be evaluated by quasi-Monte-Carlo \citep{genz2009computation}, but the special structure of (\ref{dist1}) admits a substantial reduction.

\subsection{Reduction via a shared-control latent decomposition}

Because the covariance matrix in (\ref{dist1}) has the Kronecker form $\Sigma_{\text{look}}\otimes\Sigma_{\text{arm}}$ with $\Sigma_{\text{arm}}=\bigl(\begin{smallmatrix}1&\rho\\ \rho&1\end{smallmatrix}\bigr)$ and $\Sigma_{\text{look}}=\bigl(\begin{smallmatrix}1&r\\ r&1\end{smallmatrix}\bigr)$, the dependence between the two treatment arms is entirely attributable to the shared control arm. For $\rho<1$, write
\begin{equation}\label{eq16}
Z_{k,1,i} = \sqrt{\rho}\,A_k + \sqrt{1-\rho}\,B_{k,i}, \qquad k\in\{IA,FA\},\ i=1,2,
\end{equation}
where $(A_{IA},A_{FA})\sim N(0,\Sigma_{\text{look}})$ represents the shared control arm, the pairs $(B_{IA,i},B_{FA,i})$ ($i=1,2$) are independent copies of $N(0,\Sigma_{\text{look}})$ representing the two treatment arms, and $A$ and $B$ are mutually independent. Direct calculation shows that (\ref{eq16}) reproduces (\ref{dist1}) exactly. Conditionally on $A=(a_{IA},a_{FA})$, the two treatment arms are independent, so each quadrivariate probability in (\ref{eq15}) factors into a product of two bivariate normal probabilities. Writing $\tau_k(x)=(x-\sqrt{\rho}\,a_k)/\sqrt{1-\rho}$ and, for the arguments appearing in (\ref{eq14}),
\[
\tilde A = \Phi_2\bigl(\tau_{IA}(u_1),\tau_{FA}(v_1);r\bigr),\qquad \tilde B = \Phi_2\bigl(\tau_{IA}(u_1),\tau_{FA}(v_2);r\bigr),
\]
\[
\tilde C = \Phi_2\bigl(\tau_{IA}(u_2),\tau_{FA}(v_1);r\bigr),\qquad \tilde D = \Phi_2\bigl(\tau_{IA}(u_2),\tau_{FA}(v_2);r\bigr),
\]
where $\Phi_2(\cdot,\cdot;r)$ is the standard bivariate normal CDF with correlation $r$, the five terms of (\ref{eq14}) become $\mathbb{E}_A\!\left[2\tilde A\tilde D + 2\tilde B\tilde C - 2\tilde A\tilde C - 2\tilde A\tilde B + \tilde A^2\right]$. Setting $\delta_B=\tilde B-\tilde A$, $\delta_C=\tilde C-\tilde A$, $\delta_D=\tilde D-\tilde A$ (all non-negative because $u_2\geq u_1$ and $v_2\geq v_1$), the bracket collapses algebraically to
\begin{equation}\label{eq17}
2\tilde A\tilde D + 2\tilde B\tilde C - 2\tilde A\tilde C - 2\tilde A\tilde B + \tilde A^2 \;\equiv\; \tilde A^2 + 2\tilde A\delta_D + 2\delta_B\delta_C,
\end{equation}
so that (\ref{eq15}) becomes
\begin{equation}\label{eq18}
\alpha(\eta,\gamma) = 1 - \mathbb{E}_{\zeta}\,\mathbb{E}_{A}\!\left[\tilde A^2 + 2\tilde A\delta_D + 2\delta_B\delta_C\right].
\end{equation}

\subsection{Deterministic quadrature}

The right side of (\ref{eq18}) requires only four bivariate normal probabilities per $(\zeta,A)$ node in place of five quadrivariate ones, and a bivariate normal probability admits the exact one-dimensional representation $\Phi_2(h,k;r)=\Phi(h)\Phi(k)+\int_0^r\phi_2(h,k;\varrho)\,d\varrho$ \citep{plackett1954reduction}, where $\phi_2$ is the corresponding density. Both the outer expectation over $\zeta$ and the inner expectation over $A$ can therefore be taken by Gauss--Hermite quadrature and the bivariate probabilities by Gauss--Legendre quadrature, giving a deterministic and reproducible evaluation of $\alpha(\eta,\gamma)$ with no Monte-Carlo error at any stage. In our implementation $80^2$ nodes for $\zeta$, $12^2$ nodes for $A$ and $30$ Gauss--Legendre nodes evaluate $\alpha(\eta,\gamma)$ to a tolerance of $5\times 10^{-8}$, stable under further refinement of the node counts, at every weight combination reported in the paper.

\section{A recursion for the intersection-hypothesis type 1 error at \texorpdfstring{$L$}{L} looks}\label{app:Llook}

The reduction of Appendix~\ref{app:alphaint} is confined to two looks.
Attempts to extend (\ref{eq18}) cannot survive the computational burden:
$\mathbb{E}_{\zeta}$ and $\mathbb{E}_{A}$ are each $L$-dimensional and the inner CDF is $L$-variate, so the node count is $n_{\zeta}^{L}n_{A}^{L}$ and every node carries $2^{L}$ CDF evaluations, the four of (\ref{eq18}) being the case $L=2$. The $9.2\times10^{5}$ nodes of the implemented $L=2$ setting become $8.9\times10^{8}$ at $L=3$, where the inner $\Phi_{3}$ admits no Plackett reduction and would have to be taken by quasi-Monte-Carlo, forfeiting the determinism available at $L=2$.

The remedy is a different ordering of the same integral: recurse over looks instead of integrating over them simultaneously, which is described in this appendix.
We will show that the work is $O(L\,n^{4})$: extending to $L$ looks means running the same three-dimensional loop $L$ times,
with no new dimension and no new formula.

\subsection{Setting}

Index the looks $k=1,\dots,L$, so that $k=1,2$ are the IA and FA of Section~\ref{sec:setup} when $L=2$. The statistics are those of Section~\ref{sec:setup} unchanged: $Z_{k,1,i}$ for arm $i$ versus control in Stage~1 and $\zeta_{k}$ in Stage~2, each standard normal at every look. Write $I^{(1)}_{k}$ and $I^{(2)}_{k}$ for the Stage-1 and Stage-2 information in an arm-versus-control comparison at look $k$, each proportional to the corresponding event count and increasing in $k$ under patient-wise separation, and put $\lambda^{(1)}_{jk}=\sqrt{I^{(1)}_{j}/I^{(1)}_{k}}$, $\lambda^{(2)}_{jk}=\sqrt{I^{(2)}_{j}/I^{(2)}_{k}}$ for $j\leq k$, abbreviating the one-step ratios as $\lambda^{(1)}_{k}=\lambda^{(1)}_{k-1,k}$ and $\lambda^{(2)}_{k}=\lambda^{(2)}_{k-1,k}$. Under the intersection null the two stages are independent and each carries the canonical group-sequential covariance,
\begin{equation}\label{eq:cov}
\mathrm{Cov}(Z_{j,1,i},Z_{k,1,i})=\lambda^{(1)}_{jk},\qquad
\mathrm{Cov}(Z_{j,1,1},Z_{k,1,2})=\rho\,\lambda^{(1)}_{jk},\qquad
\mathrm{Cov}(\zeta_{j},\zeta_{k})=\lambda^{(2)}_{jk},
\end{equation}
so information enters only through these ratios, and the arm-versus-arm correlation $\rho$ does not depend on $k$---which is what makes Proposition~\ref{prop:rot} available. Two features of (\ref{eq:cov}) are what the recursion uses. First, $\lambda_{jk}$ telescopes, $\lambda_{jk}=\lambda_{j+1}\lambda_{j+2}\cdots\lambda_{k}$, so each sequence is Markov, with $Z_{k,1,i}\mid Z_{k-1,1,i}=z\sim N(\lambda^{(1)}_{k}z,\,1-\lambda^{(1)2}_{k})$ and likewise for $\zeta$: the standardized statistics form an autoregression rather than a random walk, and it is the autoregression that the recursion propagates. Second, every marginal variance is $1$ whatever $k$ is, so one grid is correctly scaled at every look. At $L=2$, $\lambda^{(1)}_{2}$ is the correlation $r$ of (\ref{dist1}) and $\lambda^{(2)}_{2}$ the Stage-2 covariance in $f(\zeta_{IA},\zeta_{FA})$.

Writing $T_{k}=w_{k,1}H_{k}+w_{k,2}\zeta_{k}$ for the intersection-hypothesis statistic (\ref{eq:combined}) at look $k$, with $H_{k}=\Phi^{-1}(p^{HB}_{k,1})$, the quantity to be computed is
\begin{equation}\label{eq:Lstat}
\alpha^{\mathrm{int}}(C_{1},\dots,C_{L})=1-Pr\Bigl(\bigcap_{k=1}^{L}\{T_{k}>C_{k}\}\Bigr),
\end{equation}
where $C_{k}$ is the look-$k$ boundary and, as in Lemma~\ref{lem:key}, rejection occurs at the first $k$ with $T_{k}\leq C_{k}$. 

\subsection{The recursion}
Let $Z^{\min}_{k}=\min_{i}Z_{k,1,i}$, $Z^{\max}_{k}=\max_{i}Z_{k,1,i}$ and
\begin{equation}\label{eq:Ddouble}
D(z)=\Phi^{-1}\bigl(2\Phi(z)\bigr)\qquad(=+\infty\ \text{for}\ z\geq0),
\end{equation}
so that
\begin{equation}\label{eq:r1}
H_{k}=\min\bigl\{D(Z^{\min}_{k}),\ Z^{\max}_{k}\bigr\}.
\end{equation}
Next, $T_{k}$ is affine and increasing in $\zeta_{k}$ by (\ref{eq:combined}), so for $w_{k,2}>0$ the continuation event can be solved for $\zeta_{k}$ outright,
\begin{equation}\label{eq:r2}
\{T_{k}>C_{k}\}=\{\zeta_{k}>t_{k}\},\qquad
t_{k}=\frac{C_{k}-w_{k,1}H_{k}}{w_{k,2}},
\end{equation}
where $t_{k}$ is a function of the Stage-1 statistics alone, $H_{k}$ being evaluated from $Z_{k,1,1},Z_{k,1,2}$ by (\ref{eq:r1}). 
Next we show that one fixed rotation makes the two Stage-1 sequences independent, so that the transition density of the state factorizes across its three coordinates.

\begin{proposition}[one fixed rotation diagonalizes every transition]\label{prop:rot}
Let $P_{k}=(Z_{k,1,1}+Z_{k,1,2})/\sqrt2$ and $Q_{k}=(Z_{k,1,1}-Z_{k,1,2})/\sqrt2$. Then $P$ and $Q$ are independent Gaussian Markov sequences with $\mathrm{Var}(P_{k})=1+\rho$ and $\mathrm{Var}(Q_{k})=1-\rho$ at every look, and the state $V_{k}=(P_{k},Q_{k},\zeta_{k})$, whose values we write $v=(p,q,y)$, has the product transition density
\begin{equation}\label{eq:kern}
\kappa_{k}(v'\mid v)=\varphi_{\sigma^{P}_{k}}(p'-\lambda^{(1)}_{k}p)\,\varphi_{\sigma^{Q}_{k}}(q'-\lambda^{(1)}_{k}q)\,\varphi_{\sigma^{\zeta}_{k}}(y'-\lambda^{(2)}_{k}y),
\end{equation}
where $\varphi_{\sigma}$ is the $N(0,\sigma^{2})$ density, $\sigma^{P}_{k}=\sqrt{(1+\rho)(1-\lambda^{(1)2}_{k})}$, $\sigma^{Q}_{k}=\sqrt{(1-\rho)(1-\lambda^{(1)2}_{k})}$ and $\sigma^{\zeta}_{k}=\sqrt{1-\lambda^{(2)2}_{k}}$.
\end{proposition}
\begin{proof}
By (\ref{eq:cov}) the covariance of the Stage-1 pair between looks $j\leq k$ is $\lambda^{(1)}_{jk}\bigl(\begin{smallmatrix}1&\rho\\ \rho&1\end{smallmatrix}\bigr)$, a Kronecker product of a look factor with an arm factor, and the rotation acts on the arm factor alone: its eigenvectors are $(1,1)/\sqrt2$ and $(1,-1)/\sqrt2$ with eigenvalues $1+\rho$ and $1-\rho$, so $P$ and $Q$ are jointly Gaussian and uncorrelated at every pair of looks, hence independent, and both are independent of $\zeta$. Each therefore inherits the look factor's correlations $\lambda^{(1)}_{jk}$, which telescope, so each is Markov with a one-step law centred at $\lambda^{(1)}_{k}$ times its predecessor and variance $(1\pm\rho)(1-\lambda^{(1)2}_{k})$; that law, with the corresponding law for $\zeta$, is (\ref{eq:kern}). The eigenvectors do not depend on the look factor, and $\rho$ does not depend on $k$, so the same rotation serves at every look.
\end{proof}
In the state coordinates this makes $t_{k}$ a function $t_{k}(p,q)$ of the first two alone, since inverting Proposition~\ref{prop:rot}'s rotation gives $Z_{k,1,1}=(p+q)/\sqrt2$ and $Z_{k,1,2}=(p-q)/\sqrt2$.\\

The probability in (\ref{eq:Lstat}) is an integral over $\mathbb{R}^{3L}$; that it nevertheless fits on a three-dimensional grid is a statement about how its integrand factorizes. Write $\mathbf{1}_{k}$ for the look-$k$ continuation indicator $\mathbf{1}\{y>t_{k}(p,q)\}$, with $t_{k}$ from (\ref{eq:r2}), and
$f_{0}(v)=\varphi_{\sqrt{1+\rho}}(p)\,\varphi_{\sqrt{1-\rho}}(q)\,\varphi_{1}(y)$ for the density of $V_{1}$---the same three densities at every look, since by (\ref{eq:cov}) the marginals do not change. That integrand is a chain,
\begin{equation}\label{eq:chain}
F(v_{1},\dots,v_{L})=f_{0}(v_{1})\,\mathbf{1}_{1}(v_{1})\prod_{k=2}^{L}\kappa_{k}(v_{k}\mid v_{k-1})\,\mathbf{1}_{k}(v_{k}),
\end{equation}
in which $v_{j}$ and $v_{k}$ never appear in the same factor when $|j-k|>1$, 
because transitions are Markov, by Proposition~\ref{prop:rot},
and because each constraint is local, by (\ref{eq:r2}): the boundary is a cut on $T_{k}$, and $T_{k}$ is a function of $V_{k}$ alone.

The integrations can therefore be nested so that each eliminates a variable for good. Let $f_{k}$ be what is left of the first $k$ factors of (\ref{eq:chain}) once $v_{1},\dots,v_{k-1}$ have been integrated out---a function of one state vector, whatever $k$ is:
\begin{align}
f_{1}(v)&=f_{0}(v)\,\mathbf{1}_{1}(v),\label{eq:f1}\\
f_{k}(v')&=\mathbf{1}_{k}(v')\int f_{k-1}(v)\,\kappa_{k}(v'\mid v)\,dv,\qquad k=2,\dots,L.\label{eq:fk}
\end{align}
Equation (\ref{eq:fk}) is \emph{propagate, then cut}---the integral moves the surviving mass from look $k-1$ to look $k$, and the indicator applies the new cut---and at no point are more than three variables held. Its total mass $\pi_{k}=\int f_{k}$ is the probability of surviving looks $1,\dots,k$; the events being nested, $1=\pi_{0}\geq\pi_{1}\geq\dots\geq\pi_{L}$, and $\pi_{L}=\int F$ is the intersection probability of (\ref{eq:Lstat}).

In the canonical one-dimensional case one recurses on the test statistic itself, which is Markov and on which the boundary is a cut; here the boundary is a cut on $T_{k}$, a lumped function of $V_{k}$, and knowing $T_{k-1}$ does not determine the law of $T_{k}$. However, recursing on $V_{k}$ restores the Markov property, and $T_{k}$ is recovered pointwise from the state by (\ref{eq:r2}). Had we recursed on $T_{k}$ itself the chain (\ref{eq:chain}) would acquire long-range links, and no amount of numerical care would keep the live dimension at three. None of this is a new device: an $L$-look type 1 error is an $L$-fold integral, and the recursion of \citet{armitage1969repeated} evaluates it on a \emph{one}-dimensional grid swept $L$ times. What follows is that same calculation, with the state dimension forced up from one to three.

\subsection{Algorithm and cost}

Every integral in what follows is evaluated by the same rule: a rectangle rule on the Cartesian product of the three grids, each node taken at the centre of a cell and weighted by that cell's width, written $\Delta_{p}$, $\Delta_{q}$ and $\Delta_{y}$. The transition integral of (\ref{eq:fk}) and the masses $\pi_{k}$ are both that rule, and uniform weights are what let each kernel be a matrix and each propagation step a matrix product.

The grid is uniform rather than a Gauss rule for two further reasons. The ramp (\ref{eq:ramp}) needs cells: its partial-cell weight is exact, and is a single subtraction, only because the $y$ cells are intervals of one known width, and on Gauss nodes there is no cell and no analogue. And $f_{k}$ is not analytic---(\ref{eq:fk}) multiplies it by an indicator, so it has a kink along $y=t_{k}(p,q)$, and $t_{k}$ inherits the kinks of $H_{k}$ through (\ref{eq:r1})---so the analyticity that Gauss rules are paid for is absent exactly where the error lives.

Given $L$, the information sequences $I^{(1)}_{1}<\dots<I^{(1)}_{L}$ and $I^{(2)}_{1}<\dots<I^{(2)}_{L}$, weights with $w_{k,2}>0$, boundaries $C_{1},\dots,C_{L}$, the allocation correlation $\rho$, a grid size $n$ per coordinate and a half-width $m$ (default $7$):

\begin{enumerate}
\item[(S1)] \emph{Grids.} Take $n$ equispaced points spanning $\pm m$ standard deviations in each coordinate: $\pm m\sqrt{1+\rho}$ in $p$, $\pm m\sqrt{1-\rho}$ in $q$, $\pm m$ in $y$. By (\ref{eq:cov}) these variances do not depend on $k$, so one set serves every look at the same resolution.
\item[(S2)] \emph{Initialise.} $\pi_{0}=1$ and $f$ the product of the three look-1 normal densities.
\item[(S3)] \emph{Propagate} (looks $k\geq2$). By (\ref{eq:kern}) the transition is a product of three one-dimensional kernels, one per coordinate; acting on different axes they commute, so the step from $k-1$ to $k$ may be taken one coordinate at a time. Starting from $f^{(0)}=f_{k-1}$,
\begin{align*}
f^{(1)}(p',q,y)&=\sum_{p}\varphi_{\sigma^{P}_{k}}(p'-\lambda^{(1)}_{k}p)\,f^{(0)}(p,q,y)\,\Delta_{p},\\
f^{(2)}(p',q',y)&=\sum_{q}\varphi_{\sigma^{Q}_{k}}(q'-\lambda^{(1)}_{k}q)\,f^{(1)}(p',q,y)\,\Delta_{q},\\
f^{(3)}(p',q',y')&=\sum_{y}\varphi_{\sigma^{\zeta}_{k}}(y'-\lambda^{(2)}_{k}y)\,f^{(2)}(p',q',y)\,\Delta_{y},
\end{align*}
Each pass carries one coordinate to look $k$ and leaves the other two at look $k-1$---the primes accumulate down the display---so the three compose into the integral of (\ref{eq:fk}), the indicator still outstanding; the order in which they are taken is immaterial. Each refreshes $n^{3}$ values from an $n$-term sum, so a pass costs $n^{4}$ and a look $3n^{4}$. At $k=L$ the third pass is omitted, $f^{(2)}$ then being the $\bar f_{L-1}$ of (\ref{eq:last}), which (S5) handles instead.
\item[(S4)] \emph{Threshold surface and cut weights.} Evaluate $H_{k}$ by (\ref{eq:r1}) from $Z_{k,1,1}=(p+q)/\sqrt2$ and $Z_{k,1,2}=(p-q)/\sqrt2$, the inverse of Proposition~\ref{prop:rot}'s rotation, then $t_{k}$ by (\ref{eq:r2}). On this scale the map from $(p,q)$ to $H_{k}$ carries no $k$, so one $n\times n$ surface serves every look and only the affine step $H_{k}\mapsto t_{k}$ is redone; either way the work is $O(n^{2})$, in closed form, with no root-finding. The cut itself then follows from $t_{k}$ without discretizing it. Because (\ref{eq:r2}) makes the continuation region $\{y>t_{k}(p,q)\}$, the fraction of the cell centred at the $y$ node $y_{c}$, of width $\Delta_{y}$, that lies inside it is available exactly,
\begin{equation}\label{eq:ramp}
\omega_{c}(p,q)=\min\Bigl\{\max\Bigl\{\frac{(y_{c}+\Delta_{y}/2)-t_{k}(p,q)}{\Delta_{y}},\ 0\Bigr\},\ 1\Bigr\},
\end{equation}
a ramp rather than a step: cells lying wholly inside the region take weight $1$ and cells wholly outside take $0$, while the one cell in each $(p,q)$ column that straddles the boundary takes the exact fraction, at the cost of a subtraction.
\item[(S5)] \emph{Cut and accumulate.} For $k<L$, multiply $f$ by the ramp (\ref{eq:ramp}) and set $\pi_{k}=\int f$. At $k=L$ the Stage-2 coordinate need not be discretized at all: $\zeta_{L}$ depends on the past only through $\zeta_{L-1}$, with $\zeta_{L}\mid\zeta_{L-1}=y\sim N(\lambda^{(2)}_{L}y,\,\sigma^{\zeta2}_{L})$, and $\{T_{L}>C_{L}\}=\{\zeta_{L}>t_{L}\}$, so
\begin{equation}\label{eq:last}
\pi_{L}=\int \bar f_{L-1}(p,q,y)\;\Phi\Bigl(\frac{\lambda^{(2)}_{L}y-t_{L}(p,q)}{\sigma^{\zeta}_{L}}\Bigr)dp\,dq\,dy,
\end{equation}
where $\bar f_{L-1}$ is $f_{L-1}$ propagated in $p$ and $q$ only, and $f$ is not updated.
\item[(S6)] \emph{Return} $\alpha^{\mathrm{int}}=1-\pi_{L}$ and the exit probabilities $\pi_{k-1}-\pi_{k}$. Optionally run at two grid sizes and extrapolate in $1/n^{2}$, which buys roughly one further decimal.
\end{enumerate}

\noindent
Steps (S1)--(S2) run once and only the loop over $k$ repeats, so the cost is $O(n^{3})$ in storage---one array, overwritten in place, not growing with $L$ at all---and $O(L\,n^{4})$ in work, dominated by (S3). Calibration is a separate problem: if the boundaries are to be set by a spending function, do so on the individual-hypothesis covariance, which is exactly multivariate normal, and then run (S1)--(S6) once to read off $\alpha^{\mathrm{int}}$; only spending against the intersection test itself places the recursion inside a root-find on $C_{k}$. Replacing $H_{k}$ by $Z_{k,1,i^{*}}$ in (S4) turns the same code into the individual-hypothesis calculation, whose answer is available in closed form from an $L$-variate normal CDF, which makes that substitution an end-to-end test of (S1)--(S6).

At $L=2$ the quadrature of Appendix~\ref{app:alphaint} remains the better tool, reaching a tolerance of $5\times10^{-8}$ in about five seconds where the recursion needs some nine seconds of extrapolated work to reach $3\times10^{-7}$. That is not an accident: (\ref{eq18}) spends its budget on Gauss--Hermite and Plackett rules, which are exponentially accurate on a smooth integrand, whereas the recursion spends it on a midpoint rule through the kink of $H_{k}$, which is second order. The recursion's advantage is not accuracy per second at fixed $L$; it is that its work is linear in $L$ where that of (\ref{eq18}) is geometric, and the crossover is immediate. At $L=3$ the tensor grid grows by a factor of $n_{\zeta}n_{A}=960$, so (\ref{eq18}) would need upwards of an hour even if the inner CDF stayed as cheap---and it does not---where the recursion takes about a second.

\bibliographystyle{unsrtnat}
\bibliography{references}  






\end{document}